\documentclass[sigconf]{acmart}

\AtBeginDocument{%
  }

\copyrightyear{2024}
\acmYear{2024}
\setcopyright{acmlicensed}\acmConference[ICSE '24]{2024 IEEE/ACM 46th International Conference on Software Engineering}{April 14--20, 2024}{Lisbon, Portugal}
\acmBooktitle{2024 IEEE/ACM 46th International Conference on Software Engineering (ICSE '24), April 14--20, 2024, Lisbon, Portugal}
\acmDOI{10.1145/3597503.3639190}
\acmISBN{979-8-4007-0217-4/24/04}

\acmConference[ICSE 2024]{46th International Conference on Software Engineering}{April 2024}{Lisbon, Portugal}

\usepackage[ruled,vlined, linesnumbered]{algorithm2e}
\usepackage[noend]{algpseudocode}
\makeatletter
\renewcommand{\ALG@beginalgorithmic}{\footnotesize}
\makeatother
\usepackage{eqnarray}
\usepackage{alltt}
\usepackage{textcomp}
\usepackage{mathtools}
\usepackage{textcomp}
\usepackage{graphicx}
\usepackage{adjustbox}
\usepackage{xcolor}
\usepackage{colortbl}
\usepackage{listings}
\usepackage{xparse}

\usepackage{listings}
\usepackage{color}
\definecolor{dkgreen}{rgb}{0,0.6,0}
\definecolor{gray}{rgb}{0.5,0.5,0.5}
\definecolor{mauve}{rgb}{0.58,0,0.82}
\usepackage{pdftexcmds}
\usepackage{catchfile}
\usepackage{ifluatex}
\usepackage{ifplatform}

\iflinux
\usepackage{minted}
\fi

\usepackage{multirow}
\usepackage{macros}
\usepackage{listings,stmaryrd,array}
\usepackage{tablefootnote}
\usepackage{float}
\usepackage{enumitem}

\PassOptionsToPackage{hyphens}{url}\usepackage{hyperref}
\usepackage{xurl}

\usepackage{makecell}
\usepackage{pifont}
\newcommand{\cmark}{\ding{51}}%
\newcommand{\xmark}{\ding{55}}%

\newcommand*\justify{%
  \fontdimen2\font=0.4em
  \fontdimen3\font=0.2em
  \fontdimen4\font=0.1em
  \fontdimen7\font=0.1em
  \hyphenchar\font=`\-
}

\usepackage[normalem]{ulem}

\newcommand{\revise}[1]{{#1}}

\begin{document}

\title[]{FlashSyn: Flash Loan Attack Synthesis via Counter Example Driven Approximation (Extended Version)}         

\author{Zhiyang Chen}
\affiliation{
  \institution{University of Toronto}            
  \city{Toronto}
  \state{Ontario}
  \country{Canada}                    
}
\email{zhiychen@cs.toronto.edu}          

\author{Sidi Mohamed Beillahi}
\affiliation{
  \institution{University of Toronto}           
  \city{Toronto}
  \state{Ontario}
  \country{Canada}                    
}
\email{sm.beillahi@utoronto.ca}          

\author{Fan Long}
\affiliation{
  \institution{University of Toronto}           
  \city{Toronto}
  \state{Ontario}
  \country{Canada}                    
}
\email{fanl@cs.toronto.edu}          


\newcommand*{\name}{{FlashSyn}}
\newcommand*{\gen}{{FlashFind}}

\begin{abstract} 
    In decentralized finance (DeFi), lenders can offer flash loans to
    borrowers, i.e., loans that are only valid within a blockchain transaction
    and must be repaid with fees by the end of that transaction. Unlike
    normal loans, flash loans allow borrowers to borrow large assets
    without upfront collaterals deposits. Malicious adversaries use flash loans
    to gather large assets to exploit vulnerable DeFi protocols. 
  
  In this paper, we introduce a new framework for automated synthesis of
    adversarial transactions that exploit DeFi protocols using flash loans. To
    bypass the complexity of a DeFi protocol, we propose a new technique to
    approximate the DeFi protocol functional behaviors using numerical methods 
    (polynomial linear regression and nearest-neighbor interpolation). 
    We then construct an optimization query using the approximated functions 
    of the DeFi protocol to find an adversarial attack constituted of a sequence of 
    functions invocations with optimal parameters that gives the maximum profit. 
    To improve the accuracy of the approximation, we propose a novel
    counterexample driven approximation refinement technique. 
    We implement our framework in a tool named {\name}.
  
    We evaluate {\name} on $16$ DeFi protocols that were
    victims to flash loan attacks and
    $2$ DeFi protocols from Damn Vulnerable DeFi
    challenges. {\name} automatically synthesizes an adversarial attack for
    $16$ of the $18$ benchmarks. Among the $16$ successful cases, {\name} identifies
    attack vectors yielding higher profits than those employed by historical 
    hackers in $3$ cases, and also discovers multiple distinct attack vectors in $10$ cases, 
    demonstrating its effectiveness in finding possible flash loan attacks. 
\end{abstract}


%





\maketitle


\section{Introduction}
\label{sec:intro}

Blockchain technology enables the creation of decentralized, resilient, and
programmable ledgers on a global scale. Smart contracts, which can be deployed
onto a blockchain, allow developers to encode intricate transaction rules that
govern the ledger. These features have made blockchains and smart contracts
essential infrastructure for a variety of decentralized financial services
(DeFi). As of April 1st, 2023, the Total Value Locked (TVL) in 1,417 DeFi smart
contracts had reached $50.15$ billion~\cite{DeFillama}.

However, security attacks are critical threats to smart contracts. Attackers
can exploit vulnerabilities in smart contracts by sending malicious
transactions, potentially stealing millions of dollars from users.
Particularly, a new type of security threat has emerged, exploiting design
flaws in DeFi contracts by leveraging large amounts of digital assets. These
attacks, commonly referred to as \emph{flash loan attacks}~\cite{SlowMistStats,
mckay2022defi, zhang2023demystifying}, 
typically involve borrowing the required large amount of assets
from flash loan contracts. Among the top $200$ costliest attacks recorded in
Rekt Database,  the financial loss caused by $36$ flash loan attacks exceeds
$418$ million USD~\cite{mckay2022defi}. 


A typical flash loan attack transaction consists of a sequence of actions, or
function calls to smart contracts. The first action involves borrowing a
substantial amount of digital assets from a flash loan contract, while the last
action returns these borrowed assets. The sequence of actions in the middle
interacts with multiple DeFi contracts, using the borrowed assets to exploit
their design flaws. When a DeFi contract fails to consider corner cases created
by the large volume of the borrowed assets, the attacker may extract
prohibitive profits. For example, many flash loan attacks use borrowed assets
to temporarily manipulate asset prices in a DeFi contract to trick the contract
to make unfavorable trades with the
attacker~\cite{qin2021attacking,cao2021flashot}. Although researchers have
developed many automated program analysis and verification
techniques~\cite{mossberg2019manticore,feng2019precise, grieco2020echidna,
brent2020ethainter} to detect and eliminate bugs in smart contracts, these
techniques cannot handle flash loan attack vulnerabilities. This is because
such vulnerabilities are design flaws rather than implementation bugs.
Moreover, these techniques typically operate with one contract at a time, but
flash loan attacks almost always involve multiple DeFi contracts interacting
with each other.

\noindent \textbf{\name:}
We present {\name}, the first automated end-to-end program synthesis tool for
detecting flash loan attack vulnerabilities. Given a set of smart contracts and
candidate actions in these contracts, {\name} automatically synthesizes an
action sequence along with all action parameters to interact with the contracts
to exploit potential flash loan vulnerabilities. 
Additionally, {\name} can analyze past blockchain transaction history,
assisting users in identifying candidate actions for synthesis.

The primary challenge {\name} faces is that the underlying logic of DeFi actions
is often too sophisticated for standard solvers to handle. Even
if the action sequence was already known, a naive application of symbolic
execution might not be able to find action parameters because it may need to extract
overly complicated symbolic constraints causing the solvers to
time out. Moreover, {\name} synthesizes the action sequence and the action
parameters together and therefore faces an additional search space explosion
challenge.

{\name} addresses these challenges with its novel
\emph{synthesis-via-approximation} technique. Instead of attempting to extract
accurate symbolic expressions from smart contract code, {\name} collects data
points to approximate the effect of contract functions with numerical methods. {\name} then uses the approximated expressions to drive the
synthesis. {\name} also incrementally improves the approximation with its novel
\emph{counterexample driven approximation refinement} techniques, i.e., if the
synthesis fails because of a large deviation caused by the approximations,
{\name} collects the corresponding data points as counterexamples to
iteratively refine the approximations. The combination of these techniques
allows the underlying optimizer of {\name} to work with more tractable
expressions. It also decouples the two difficult tasks, finding the action
sequence and finding the action parameters. When working with a set of
coarse-grained approximated expressions, {\name} can filter out unproductive
action sequences with a small cost.

\noindent \textbf{Experimental Results:} 
We evaluate {\name} on $16$ DeFi benchmark protocols that were victims to flash
loan attacks and $2$ DeFi benchmark protocols from Damn Vulnerable DeFi
challenges~\cite{DVD}. {\name} synthesizes adversarial attacks for $16$ out
of the $18$ benchmarks. For comparison, a baseline with manually crafted
accurate action summaries only synthesizes attacks for $7$ out of the $18$.

\noindent \textbf{Contributions:}
This paper makes the following contributions:
\begin{itemize}[leftmargin=*]
\item \textbf{\name:} The first automated
end-to-end program synthesis tool for detecting flash loan attack
vulnerabilities. It enables approximate attack synthesis 
without diving into sophisticated logics of DeFi contracts. 
\item \textbf{Synthesis-via-approximation:} A novel \revise{synthesis-via-approxi- mation} technique to handle sophisticated logics of DeFi contracts. 
\item \textbf{Counterexample Driven Approximation Refinement:} A novel counterexample driven approximation refinement technique to
        incrementally improve the approximation during the synthesis process.
\item \textbf{Experimental Evaluation:} 
We implemented {\name} in a tool and evaluated it on $16$ protocols that were victims to flash
loan attacks and $2$ fictional flash loan attacks. 
\end{itemize}

Our solution, {\name} has been adopted and further developed by Quantstamp, a leading smart contract auditing company for the detection of flash loan vulnerabilities in DeFi contracts~\cite{adoption,adoptionNews1,adoptionNews2}. 
\section{Background}
\label{sec:background}

\noindent \textbf{Blockchain:}
Blockchain is a distributed ledger that broadcasts and stores information of
transactions across different parties. 
A blockchain consists of a growing
number of blocks and a consensus algorithm determining block order.
Each block is constituted of transactions.
Ethereum~\cite{ethereum22,buterin2014ethereum} is the first blockchain to
support, store, and execute Turing complete programs, known as smart contracts.
Many new blockchains use the Ethereum virtual machine (EVM) 
for execution due to its popularity among developers.


\noindent \textbf{Smart Contracts:}
Each smart contract is associated with a unique address, a
persistent account's storage trie, a balance of native tokens, e.g., Ether in
Ethereum, and bytecode (e.g., EVM
bytecode~\cite{ethereum22,buterin2014ethereum}) that executes incoming
transactions to change the storage and balance. Users interact with a smart
contract by issuing transactions from their user accounts to the contract
address. 
Smart contracts can also interact with other smart contracts as function calls.
Currently, there are several human-readable high-level programming languages,
e.g., Solidity~\cite{solidity} and Vyper~\cite{vyper}, to write smart contracts
that compile to the EVM bytecode.

\noindent \textbf{Decentralized Finance (DeFi):}
DeFi is a peer-to-peer financial ecosystem built on top of
blockchains~\cite{wust2018you}. The building blocks of DeFi are smart contracts
that manage digital assets. 
A few DeFi protocols dominate the DeFi market and serve as references for other
decentralized applications: stable coins (e.g., USDC and USDT), price oracles,
decentralized exchanges, and lending and borrowing platforms. 
In DeFi, a special type of loan called \textbf{flash
loan} allows lenders to offer loans to borrowers without upfront collaterals
deposits. The loan is only valid within a single transaction and must be repaid
with fees before the completion of the transaction.



\section{Illustrative Example}
\label{sec:example}

We next present a motivating example to describe the complexity of flash loan attacks and our proposed approach to synthesize them.

\begin{figure}[]
    \centering
    \includegraphics[width=0.9\linewidth]{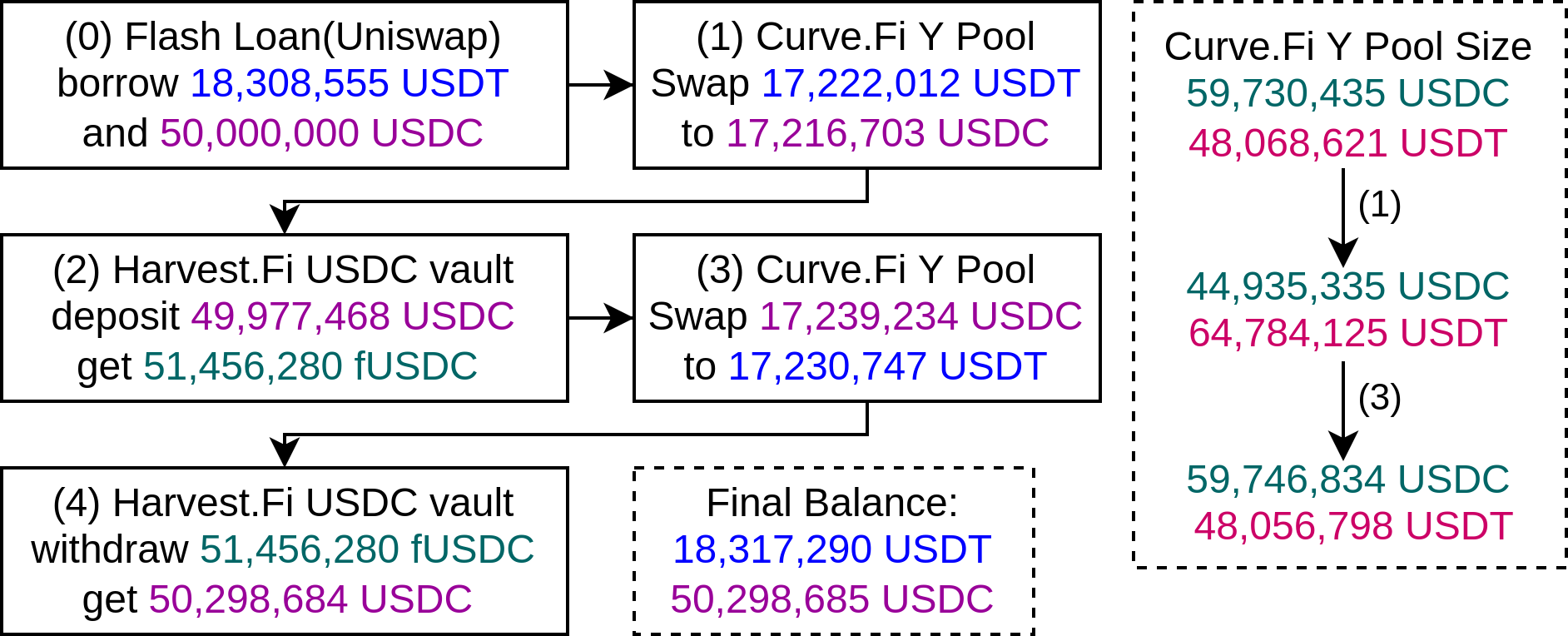}
    \caption{Harvest USDC Vault Price Manipulation Attack.}
    \label{fig:Harvest_USDC}
    \vspace*{-2mm}
\end{figure}

\noindent \textbf{Background:} On October 26th 2020, an attacker exploited the
USDC and USDT vaults of Harvest Finance, causing a financial loss of about
$33.8$ million USD. In this section, we will focus on the attack on the USDC vault.
The attacker repeatedly executed the same attack vector 17 times targeting the USDC vault.
Fig.~\ref{fig:Harvest_USDC} summarizes the attack vector.
The attack vector contains a sequence of actions that interact with the following contracts:
\begin{itemize}[leftmargin=*]
    \item \textbf{Uniswap:} Uniswap is a protocol with flash loan services. 

    \item \textbf{Curve:} Curve is an exchange protocol for stable coins like
        USDT, USDC, and DAI, whose market prices are close to one USD. It
        maintains pools of stable coins and users can interact with these pools
        to exchange one kind of stable coins to another. For example, Y Pool in
        Curve contains both USDC and USDT. Users can put USDC into the pool to
        exchange USDT out. The exchange rate fluctuates around one, which is
        determined by the current ratio of USDC and USDT in the pool. Note that
        internally Y pool automatically deposits USDT and USDC to \emph{Yearn}\revise{,}\footnote{Yearn is a DeFi protocol that generates yield on deposited assets. yTokens of Yearn represent the liquidity provided in a Yearn product.} keeps yUSDT and yUSDC tokens, and retrieves them back when the users withdraw. We omit this complication for simplicity.

    \item \textbf{Harvest:} Harvest is an asset management protocol and the
        victim contract of this attack. Users can deposit USDC and USDT into
        Harvest and receive fUSDC and fUSDT tokens which users can later use to
        retrieve their deposit back. Harvest will invest the deposited USDC and
        USDT from users to other DeFi protocols to generate profit. Note that
        the exchange rate between fUSDC and USDC is also not fixed. It is
        determined by a vulnerable closed source oracle contract, which ultimately
        uses Curve Y pool ratios to calculate the exchange rate.
\end{itemize}


\vspace*{1mm}
\noindent \textbf{Attack Actions:} The attack vector shown in
Fig.~\ref{fig:Harvest_USDC} first flash loaned $18.3$M USDT and $50$M USDC,
then called $4$ methods (actions) to exploit the design flaw in Harvest. The
first action, action $1$, swaps $17,222,012$ USDT for $17,216,703$ USDC via 
\emph{Curve.Fi Y Pool}. 
Action $1$ reduces the estimated value of USDT based on the ratio in Y pool as
the amounts of USDT in Y Pool considerably increases. This in turn reduces
Harvest Finance's evaluation of its invested assets. Action $2$ deposits
$49,977,468$ USDC into \emph{Harvest Finance USDC vault} and due to the reduced
evaluation of the invested underlying assets, the attacker receives
$51,456,280$ fUSDC back, which is abnormally large. Similar to action $1$,
action $3$ then swaps $17,239,234$ USDC back to $17,230,747$ USDT via
\emph{Curve.Fi Y Pool}, which normalized the manipulated USDT/USDC rate in the
pool. It also brings Harvest Finance's evaluation of its invested underlying
asset back to normal. Finally, action $4$ withdraws $50,298,684$ USDC (using
$51,456,280$ fUSDC) from \emph{Harvest Finance USDC vault}. Assuming $1$ USDC
$= 1$ USDT $= 1$ USD, the profit of the above attack vector is $307,420$ USD. 

This attack is a typical case of oracle manipulation. The exploiter manipulated
the USDT/USDC rate in \emph{Curve.Fi Y Pool} by swapping a large amount between
USDC and USDT back and forth, which caused Harvest Finance protocol to
incorrectly evaluate the value of its asset, leaving large arbitrage space for
the exploiter. The actions sequence and particularly the parameters are
carefully chosen by the attacker to yield best profit. There are multiple
challenges {\name} faces to synthesize this attack.


\noindent \textbf{Challenge 1 - Sophisticated Interactions:} The attack
involves several smart contracts that interact with each other and with
other contracts outside the attack vector. The state changes caused by one action
influence the behavior of other actions.
This makes the synthesis problem of finding an attack vector more complicated
as the effect of an action depends on its predecessor actions thus actions
cannot be treated separately.

\noindent \textbf{Challenge 2 - Close Source:} Some external smart contracts
that a DeFi protocol interacts with are not \emph{open-source}. For instance,
the source code of the external smart contract
\emph{PriceConverter}\footnote{Ethereum address:
0xfca4416d9def20ac5b6da8b8b322b6559770efbf.} of Harvest Finance protocol is not
available on Etherscan~\cite{etherscan}, and it is called by actions $2$ and
$4$ to determine the exchange rate between fUSDC and USDC.
This impedes the complete understanding of the DeFi protocol implementation and
to reason about its correctness to anticipate attacks vectors.

\noindent \textbf{Challenge 3 - Mathematical Complexity:} DeFi contracts use
mathematical models that are too complex to reason about. For instance, actions
$2$ and $4$ swap an amount of token $i$ to token $j$, while maintaining the
following \emph{StableSwap invariant}~\cite{egorov2019stableswap}: 

$A\cdot n^n\sum_i x_i  + D = A \cdot n^n \cdot D + \frac{D^{n+1}}{n^n \prod_{i}{x_i}}$
where $A$ is a constant, $n$ is number of token types in
the pool($4$ for \emph{Curve.Fi Y Pool})\footnote{Ethereum address: 0x45f783cce6b7ff23b2ab2d70e416cdb7d6055f51.},
 $x_i$ is token $i$'s liquidity,
$D$ is the total amount of tokens at equal prices.
There does not exist a closed-form solution for $D$ as it requires finding
roots of a \emph{quintic equation}. In the actual implementation, $D$ is calculated iteratively on the fly via Newton's method (see Appendix A).





To demonstrate the complexity of DeFi protocols, we run an experiment with
Manticore~\cite{mossberg2019manticore}, a symbolic execution tool for smart
contracts, to execute the function \emph{get\_D}, for computing \emph{D} as
shown in Fig.~\ref{fig:new_get_D}, with symbolic inputs and explore all
possible reachable states. Manticore fails and throws a solver-related
exception together with an \emph{out of memory} error. We then simplified
\emph{get\_D} by removing the outer \emph{for} loop and bounding the length of
\emph{xp} to $2$, Manticore still fails and throws the same error. 




\definecolor{verylightgray}{rgb}{.97,.97,.97}

\lstdefinelanguage{Solidity}{
	keywords=[1]{anonymous, assembly, assert, balance, break, call, callcode, case, catch, class, constant, continue, constructor, contract, debugger, default, delegatecall, delete, do, else, emit, event, experimental, export, external, false, finally, for, function, gas, if, implements, import, in, indexed, instanceof, interface, internal, is, length, library, log0, log1, log2, log3, log4, memory, calldata, modifier, new, payable, pragma, private, protected, public, pure, push, require, return, returns, revert, selfdestruct, send, solidity, storage, struct, suicide, super, switch, then, this, throw, transfer, true, try, typeof, using, value, view, while, with, addmod, ecrecover, keccak256, mulmod, ripemd160, sha256, sha3}, 
	keywordstyle=[1]\color{blue}\bfseries,
	keywords=[2]{address, bool, byte, bytes, bytes1, bytes2, bytes3, bytes4, bytes5, bytes6, bytes7, bytes8, bytes9, bytes10, bytes11, bytes12, bytes13, bytes14, bytes15, bytes16, bytes17, bytes18, bytes19, bytes20, bytes21, bytes22, bytes23, bytes24, bytes25, bytes26, bytes27, bytes28, bytes29, bytes30, bytes31, bytes32, enum, int, int8, int16, int24, int32, int40, int48, int56, int64, int72, int80, int88, int96, int104, int112, int120, int128, int136, int144, int152, int160, int168, int176, int184, int192, int200, int208, int216, int224, int232, int240, int248, int256, mapping, string, uint, uint8, uint16, uint24, uint32, uint40, uint48, uint56, uint64, uint72, uint80, uint88, uint96, uint104, uint112, uint120, uint128, uint136, uint144, uint152, uint160, uint168, uint176, uint184, uint192, uint200, uint208, uint216, uint224, uint232, uint240, uint248, uint256, var, void, ether, finney, szabo, wei, days, hours, minutes, seconds, weeks, years},	
	keywordstyle=[2]\color{teal}\bfseries,
	keywords=[3]{block, blockhash, coinbase, difficulty, gaslimit, number, timestamp, msg, data, gas, sender, sig, value, now, tx, gasprice, origin},	
	keywordstyle=[3]\color{violet}\bfseries,
	identifierstyle=\color{black},
	sensitive=false,
	comment=[l]{//},
	morecomment=[s]{/*}{*/},
	commentstyle=\color{gray}\ttfamily,
	stringstyle=\color{red}\ttfamily,
	morestring=[b]',
	morestring=[b]"
}

\lstset{
	language=Solidity,
	backgroundcolor=\color{verylightgray},
	extendedchars=true,
	basicstyle=\footnotesize\ttfamily,
	showstringspaces=false,
	showspaces=false,
	numbers=left,
	numberstyle=\footnotesize,
	numbersep=9pt,
	tabsize=2,
	breaklines=true,
	showtabs=false,
	captionpos=b
}

\begin{figure}[]
    \centering
    \begin{minipage}[h]{0.9\linewidth}
\begin{lstlisting}
function get_D(uint[] xp) returns (uint):
  uint N_COINS = xp.length;    
  uint S = sum(xp);
  // ...
  uint D = S;      
  uint Ann = A * N_COINS; // A is a constant  
  for (uint i = 0; i < 255; i = i + 1) {
    uint D_P = D;
    for (uint j = 0; j < xp.length; j = j + 1) {
      D_P = (D_P * D) / (xp[j] * N_COINS + 1); 
    }
    uint Dprev = D;
    D = ((Ann * S + D_P * N_COINS) * D) /
      ((Ann - 1) * D + (N_COINS + 1) * D_P);
      if (abs(D - Dprev) <= 1) break; 
  }
  return D;  
\end{lstlisting}
    \vspace{-7mm}
    \end{minipage}
    \caption{\emph{get\_D} Method to Compute \emph{D}.}
    \vspace*{-5mm}
\label{fig:new_get_D}
\end{figure}

\subsection{Apply \name{}}



We will now show how {\name} synthesizes the Harvest USDC vault attack from the identified set of actions listed in Table~\ref{table:prepost_harvest_usdc}.\footnote{In Section~\ref{sec:impl}, we present \gen{} to automatically find the set of candidate actions that are used here by {\name} to synthesize an attack vector.} 
The first two input arguments to \codeword{exchange} specify the token types to be swapped.
The third argument specifies the quantity to swap.\footnote{Note that in the
implementation the actual name of the \codeword{exchange} method is
\codeword{exchange\_underlying}, $1$ and $2$ are used to identify the tokens
USDC and USDT, respectively, and the method has a fourth argument to specify
the minimal quantity expected to receive from the swapping.}
Table~\ref{table:prepost_harvest_usdc} lists each action's token flow, along with the number of data points collected initially (without counterexamples) and the total number of data points for polynomial and interpolation, respectively. The amounts of tokens transferred in/out for each action are calculated based on its contract's member variables or read-only functions. We refer these variables and functions as \textbf{states} of an action. A \textbf{prestate} refers to the state before the execution of an action. Executing an action will also alter states, which are denoted as \textbf{poststates}. The states not altered by any action are ignored.  


\justify
For example, \texttt{exchange(USDT,USDC,v)} leverages two states, \texttt{balances[USDC]} and \texttt{balances[USDT]}, to calculate the amounts of token exchanges. \revise{Upon execution, this function also modifies these two states. Consequently, these two states act as both the prestates and poststates of \texttt{exchange(USDT,USDC,v)}.}


\noindent \textbf{Initial Approximation:}
To generate the initial approximation of the state transition functions of each
action, {\name} first collects data points where each data point is an \textit{input-output} pair. The \textit{input} is the
action's prestates and parameters, and the \textit{output} is its poststates
and the outputted values. 
To collect data points, {\name} executes the associated contracts on a private
blockchain (a forked blockchain environment) with different parameters to reach
\textit{input-output} pairs with different prestates and poststates.    
{\name} then uses the collected data points to find the approximated state
transition functions. 
We consider two techniques to solve the above multivariate approximation
problem: linear regression based polynomial features and nearest-neighbor
interpolation~\cite{montgomery2021introduction, rukundo2012nearest}. 
The following example is one of \codeword{exchange(USDT,USDC,v)}'s state transition functions approximated by polynomials: $\stateof{1}' = 0.73244455 \times \stateof{1} -
0.23655202 \times \stateof{2} - 0.85915531 \times v + 27351279.416023515$ where 
$\stateof{1}$ and $\stateof{1}'$ are the prestate and poststate \texttt{balances[USDC]},  
$\stateof{2}$ is the prestate \texttt{balances[USDT]}, and $v$ is the third argument of the action \codeword{exchange(USDT,USDC,v)}.

\begin{table}[]
    \caption{Actions in Harvest USDC Vault Attack. IDP and TDP denotes the initial and total number of datapoints. USDT(-) (resp., USDC(+)) denotes USDT (resp., USDC) tokens transferred out (reps., in).}
    \label{table:prepost_harvest_usdc}
    \vspace{-0.2cm}
    \centering
    \resizebox{\columnwidth}{!}{%
    \begin{tabular}{|l|l|l|l|l|}
    \hline
    Action          & Token Flow                            & IDP & TDP-Poly & TDP-Inter\\  \hline
    exchange        & \multirow{2}{*}{USDT(-), USDC(+)}     & 2000 & 2238 & 2792      \\ 
    (USDT, USDC, v) &                                       &  &  &  \\\hline
    exchange        & \multirow{2}{*}{USDC(-), USDT(+)}     & 2000 & 2148 & 2888     \\ 
    (USDC, USDT, v) &                                       &  &  &      \\ \hline
    deposit(v)      & USDC(-), fUSDC(+)                     & 2000 & 2162 & 2358   \\ \hline
    withdraw(v)     & fUSDC(-), USDC(+)                     & 2000 & 2364 & 2876      \\ \hline
    \end{tabular}
    }
    \vspace*{-7mm}
\end{table}


\noindent \textbf{Enumerate and Filter Action Sequences:} After capturing an
initial approximation of state transition functions, {\name} leverages an
enumeration-based top-down algorithm to synthesize different action sequences. 
{\name} applies several pruning heuristics to filter
unpromising sequences.
For each enumerated action sequence, {\name} uses the approximated state
transition functions to construct an optimization problem, consisting of constraints and an
objective function that represents profit. {\name} then applies
an off-the-shelf optimizer to obtain a list of parameters that maximize the
profit estimated using approximated transition functions. 



\noindent \textbf{Counterexample Driven Refinement:} 
After obtaining a list of parameters that maximize the estimated profit of an action sequence, {\name} proceeds to verify the synthesized attack vectors by executing them 
on a private blockchain and check their actual profits. If the difference between the actual profit and the
estimated profit of an attack vector is greater than $5\%$, {\name} reports it as a counterexample, indicating inaccuracy of our approximated transition functions. To correct
this inaccuracy, {\name} employs \textit{counterexample driven
approximation refinement} technique. {\name} utilizes the reported counterexamples
to collect new data points and refine the approximations. The revised approximations
are subsequently used to search for parameters in next loops. For example, {\name} with polynomial approximations collects $238$ additional data points for the action \codeword{exchange(USDT, USDC, v)} throughout $7$ refinement loops.

\noindent \textbf{Synthesized Attack:} In the Harvest USDC example, {\name}
successfully found the following attack vector that yields an adjusted profit
of $110051$ USD using the interpolation technique with the counterexample
driven refinement loop. 

\codeword{exchange(USDT, USDC, 15192122)}  $\cdot$ \codeword{deposit(45105321)}\  $\cdot$ \\
\codeword{exchange(USDC, USDT, 11995404)}  $\cdot$ \codeword{withdraw(46198643)}





\section{Preliminary}
\label{sec:preli}

\textbf{Labeled Transition Systems (LTS).} 
We use LTS to model behaviors of smart contracts. 
A LTS $A=(Q,\Sigma,  \mathsf{q}_0, \delta)$ over the possibly-infinite alphabet
$\Sigma$ is a possibly-infinite set $Q$ of states with
an initial state $\mathsf{q}_0 \in Q$, and a transition relation $\delta \subseteq Q \times \Sigma \times
Q$. 

\noindent
\textbf{Execution.} 
An \emph{execution} of $A$ is a sequence of states and transition labels (\emph{actions}) $\rho = \mathsf{q}_0, \mathsf{a}_0,\mathsf{q}_1\ldots \mathsf{a}_{k-1},\mathsf{q}_k$ for $k>0$ such that $\delta(\mathsf{q}_i, \mathsf{a}_i, \mathsf{q}_{i+1})$ for each $0\leq i<k$. We write $\mathsf{q}_i\xrightarrow{\mathsf{a}_i\ldots \mathsf{a}_{j-1}}_A \mathsf{q}_j$ to denote the subsequence $\mathsf{q}_i,\mathsf{a}_i,...,\mathsf{q}_{j-1},\mathsf{a}_{j-1},\mathsf{q}_j$ of $\rho$.

\noindent
\textbf{Invocation Label.} 
Formally, an \emph{invocation label} $\adr.m(\vec{u})$ consists of a method name $m$ of a contract address $\adr$, accompanied by a vector $\vec{u}$ containing argument values.

\noindent
\textbf{Operation Label.} 
An \emph{operation label} $\ell :=
\adr.m(\vec{u}) \Rightarrow (I,v) \ \cup\ \bot$ is an invocation label $\adr.m(\vec{u})$ along with a
return value $v$ 
, and $I$ is a sequence of operation labels representing the ``internal'' calls made during the invocation of $m$. 
The distinguished invocation outcome $\bot$ is associated to invocations that revert.

\noindent
\textbf{Interface.}
The \emph{interface} $\Sigma_{\adr}$ is the set of non-read-only operation labels in the contract $\adr$. We assume w.l.o.g. that the preconditions are satisfied for all the operations in $\Sigma_{\adr}$, otherwise, the external invocation $\adr.m(\vec{u})$ reverts. $\Sigma_{\adr}$ is a superset of the set of action candidates of {\name}.


\noindent
\textbf{Smart Contract.}
A \emph{smart contract} at an address $\adr$ is an LTS $C_{\adr}=(Q_{\adr},\Sigma_{\adr}, \mathsf{q}_0, \delta_{\adr})$ over the interface $\Sigma_{\adr}$ where $Q_{\adr}$ is the set states and $\delta_{\adr}$ is the transition relation.



\noindent
\textbf{Symbolic Actions Vector.}
We define the notion of a symbolic actions vector $\symbolic = \ell_{\adr 1} \ldots \ell_{\adr n}$ s.t.  $\ell_{\adr i}\in \Sigma$ for $1\leq i<n$ as the sequence of operation labels (possibly from different  contracts) associated with the execution $\rho$, i.e., $\rho = \mathsf{q}_1, \ell_{\adr 1},\mathsf{q}_1\ldots \ell_{\adr n},\mathsf{q}_n$.



\noindent
\textbf{Balance.} 
We define the balance of address $\adr$ in a blockchain state $\mathsf{q}$ as the mapping $\Balance: Q \times \mathbf{A} \implies \mathbf{V}$
    that maps the pair $(\mathsf{q},\adr) \in Q \times \mathbf{A}$ to the weighted sum of tokens the address $\adr$ holds at $\mathsf{q}$, i.e., 
    $\Balance(\mathsf{q},\adr)= \sum_{\mathsf{t} \in \mathbf{T}}{\tokenmap(\mathsf{q},\adr,\mathsf{t}) \cdot \tokenpricemap(\mathbf{t})}$, where $\mathbf{T}$ represents tokens hold by $\adr$, $\tokenmap(\mathsf{q},\adr,\mathsf{t})$ represents the amount of token $\mathsf{t}$ hold by $\adr$ at the blockchain state $\mathsf{q}$, and $\tokenpricemap(\mathsf{q}, \mathbf{t})$ represents the price of token $\mathsf{t}$ at the blockchain state $\mathsf{q}$.



%
%
\noindent
\textbf{Attack Vector.}
    An \emph{attack vector} by an adversary $\adr$ consists of a symbolic actions vector $\symbolic$ where the symbolic arguments are replaced by concrete values (integer values) and $\symbolic$ transforms a blockchain state $\mathsf{q}$ to another state $\mathsf{q}'$ such that $\Balance(\mathsf{q}',\adr) -  \Balance(\mathsf{q},\adr) > 0$, i.e., the adversary $\adr$ generates profit when the sequence of actions $\symbolic$ is executed with the concrete values.

\noindent
\textbf{Problem formulation.}
Given a specification $\spec$ (which contains vulnerable contract addresses or action candidates) and a blockchain state $\mathsf{q}$, the objective is to find an attack vector consisting of a concretization of the symbolic actions vector $\symbolic = \ell_{\adr 1} \ldots \ell_{\adr n}$ s.t.  $\ell_{\adr i}\in \Sigma \cap \spec$  for $1\leq i<n$, transforming the state $\mathsf{q}$ to a state $\mathsf{q}'$, and that maximizes the profit of an adversary $\adr$, $\Balance(\mathsf{q}',\adr) -  \Balance(\mathsf{q},\adr)$.



\section{\name{}}
\label{sec:approach}



\subsection{Symbolic Actions Vectors Synthesis}
Algorithm~\ref{alg:enu} gives the overall synthesis procedure of {\name}. {\name} first collects initial data points to approximate the actions in $\actionset$ (line~\ref{algo1:line1}) where {\name} uses the state $\q$ as a starting blockchain state. Then, using the sub-procedure \textsc{Approximate} {\name} generates the approximations $\approxact$ of the actions in $\actionset$ using the collected data points (line~\ref{algo1:line3}). {\name} uses the sub-procedure \textsc{ActionsVectors} to generate all possible symbolic actions vectors of length less than $\len$ (line~\ref{algo1:line4}). 
{\name} then iterates over the generated actions vectors and uses some heuristics implemented in the sub-procedure \textsc{IsFeasible} to prune actions vectors (line~\ref{algo1:line6}). For instance, an actions vector containing two adjacent actions invoking the same method
can be pruned to an actions vector where the two adjacent actions are merged. Afterwards, using the actions vector and  approximated transition functions, 
 the sub-procedure \textsc{Construct} constructs the optimization framework $\mathcal{P}$ for the actions vector (line~\ref{algo1:lineX}).
Then, {\name} uses the optimization sub-procedure \textsc{Optimize} (line~\ref{algo1:line7}) to find the optimal concrete values to pass as input parameters to the methods in the actions vector that satisfy the constraints of $\mathcal{P}$. {\name} then validate whether the attack vector generated by the optimizer indeed generates the profit with the sub-procedure \textsc{QueryOracle} to execute the generated attack vectors on the blockchain. If the query is successful, i.e., the actual profit closely matches the profit found by the optimizer, {\name} adds the attack vector to the list of discovered attacks. Otherwise, {\name} considers the attack vector to be a counterexample, and uses it to generate new data points to refine the approximation in the subsequent iterations, \revise{within the sub-procedure \textsc{CEGDC} (referenced in line~\ref{algo1:line10} and introduced later in Section~\ref{sec:cegdc})}. {\name} repeats the process until the number of iterations reaches $\iter$ (line~\ref{algo1:line2}).

\setlength{\textfloatsep}{10pt}
\begin{algorithm}[]
  \caption{Attack vectors synthesis procedure. Its inputs are actions $\actionset$, the maximum length $\len$, a blockchain state $\q$, and a threshold number of iterations $\iter$. Its outputs are attack vectors that yield positive profits.}\label{alg:enu}
  \begin{algorithmic}[1]
  \Procedure{Synthesize}{$\actionset$, $\len$, $\mathsf{P}$, $\q$, $\iter$}
  \State \ \ \textbf{for each}\ $\act \in \actionset$ \label{algo1:line0}
  \State \ \ \ \ $\mathsf{datapoints}[\act] \leftarrow \textsc{DataCollect}(\q, \act);$ \label{algo1:line1}
  \State \ \ \textbf{for each}\ $i \in [0;\ \iter]$ \label{algo1:line2}
  \State \ \ \ \ $\approxact \leftarrow \textsc{Approximate}(\actionset,\mathsf{datapoints});$ \label{algo1:line3}
  \State \ \ \ \ $\wkp \leftarrow \textsc{ActionsVectors}(\approxact, \len);$ \label{algo1:line4}
  \State \ \ \ \ \textbf{for each}\ $\pp \in \wkp$ \label{algo1:line5}
  \State \ \ \ \ \ \ \textbf{if}\ $\textsc{IsFeasible}(\pp)$ \label{algo1:line6}
  \State \ \ \ \ \ \ \ \ $\mathcal{P} \leftarrow \textsc{Construct}(\pp, \approxact)$ \label{algo1:lineX}
  \State \ \ \ \ \ \ \ \ $(\pp^\star, \textsf{profit}) \leftarrow \textsc{Optimize}(\pp, \mathcal{P})$; \label{algo1:line7}
  \State \ \ \ \ \ \ \ \ \textbf{if}\ $\textsc{QueryOracle}(\q, \pp^\star, \textsf{profit})$ \label{algo1:line8}
  \State \ \ \ \ \ \ \ \ \ \ $\al.\textsf{add}(\pp^\star,\textsf{profit})$; \label{algo1:line9}
  \State \ \ \ \ \ \ \ \ \textbf{else} 
  \State \ \ \ \ \ \ \ \ \ \ $\mathsf{datapoints} \assign \mathsf{datapoints} \cup \textsc{CEGDC}(\pp^\star, \q)$; \label{algo1:line10}
  \State \ \ \textbf{return}\ $\al$;
  \EndProcedure
  \end{algorithmic}
\end{algorithm}

\vspace{-2mm}
\subsection{Pruning Symbolic Actions Vectors} \label{sec:isfeasible} 
The sub-procedure \textsc{IsFeasible} implements some heuristics to prune undesired symbolic actions vectors. 

\noindent \textbf{Heuristic 1: no duplicate adjacent actions.} Two successive calls to the same method in a DeFi smart contract are usually equivalent to a single call with larger parameters. Thus, we discard actions vectors containing duplicate, successive actions. 

\noindent \textbf{Heuristic 2: limited usage of a single action.} Using the observation that attack vectors do not contain repetitions, we fix a maximum number of calls to a single method an attack can contain and discard actions vectors that do not satisfy this criterion, e.g., an actions vector of length $4$ cannot contain more than $2$ calls to the same method. 

\noindent \textbf{Heuristic 3: necessary preconditions.} Based on the observation that owning certain tokens is a necessary precondition for invoking some actions, {\name} prunes symbolic actions vectors that contain actions requiring tokens\footnote{Note a token can be standard tokens (ERC20, BEP20), or any other forms of tokens such as debt tokens or share tokens.} not owned by the attacker. For example, in Harvest USDC example, invoking \codeword{withdraw} method requires users own some share tokens (fUSDC) beforehand.  The only action candidate that mints fUSDC for users is \codeword{deposit}; thus, this heuristic mandates that \codeword{deposit} must be called before invoking \codeword{withdraw}. This heuristic establishes a necessary but not sufficient condition to ensure that synthesized attack vectors will not be reverted.

\vspace{-2mm}
\subsection{Optimization} \label{sec:opframe} 
Given a symbolic actions vector and their approximated transit functions, the sub-procedure \textsc{Construct} constructs an optimization framework to find optimal values for the parameters for the actions. Recall that given a blockchain state $\q$ and an address $\adr$, the actions vector $\symbolic$ transforms $\q$ to another state $\q'$. The objective function in the optimization problem targets to increase the tokens values in the balance of the address $\adr$, i.e., $y = \Balance(\q',\adr) -  \Balance(\q,\adr)$. The optimization problem is accompanied by constraints on the symbolic values to be inferred. For instance, the balance of any token $\mathbf{t}$ for any address $\adr'$ must always be non-negative, i.e., the adversary and the smart contracts cannot use more tokens than what they have in their balances, otherwise the transaction reverts. 
In the following, we give the definition of the optimization problem.

\vspace{-3mm}
\begin{align*}  
\mathcal{P}: \ \ \ & \begin{cases} 
    \max_{p_0, p_1, ..., p_n}\ y = \Balance(\q',\adr) -  \Balance(\q,\adr) \\ 
   \textrm{subject to: } \forall\ \mathbf{t} \in \mathbf{T}, \adr' \in \mathbf{A}.\ \tokenmap(\q',\adr',\mathbf{t})  \geq 0   
    \end{cases}       
\end{align*}

\subsection{Counterexample Guided Data Collection (CEGDC)} \label{sec:cegdc}
The optimization sub-procedure might explore parts of the states space not explored during the initial data points collection. This might challenge the accuracy of the approximations and result in mismatch between the estimated  and the actual values. 
Thus, it is necessary to collect new data points based on the counterexamples that show the mismatch between the estimated and the actual values, to refine the approximations. 
Therefore, we propose counterexample guided data collection (CEGDC), inspired of counterexample guided abstraction refinement~\cite{clarke2000counterexample}, to refine approximations when mismatches are identified. 

We use $\concrete$ to denote the attack vector s.t. $\q \xrightarrow{\concrete} \q'$. 
$\q'_{e}$ and $\q'_{a}$ denote the estimated value for the state $\q'$ found by the optimizer and the actual value obtained when executing $\concrete$ on the actual protocol on the blockchain, respectively. 
$\profit_e(\concrete) = \Balance(\q'_{e},\adr) -  \Balance(\q,\adr)$ and $\profit_a(\concrete) = \Balance(\q'_{a},\adr) -  \Balance(\q,\adr)$ denote the estimated profit and actual profit, respectively. 

\begin{definition} \label{counterexample}
A counterexample is an attack vector $\concrete$ whose estimated profit $\profit_e(\concrete)$ is different from its actual profit $\profit_a(\concrete)$. Formally, $|\profit_e(\concrete) - \profit_a(\concrete)| \geq \varepsilon \cdot (|\profit_e(\concrete)| + |\profit_a(\concrete)|) $, where $\varepsilon$ is a small constant representing accuracy tolerance.
\end{definition}

\vspace*{-5mm}
\setlength{\textfloatsep}{10pt}
\begin{algorithm}[]
  \caption{Counterexample guided data collection procedure. It takes a counterexample $\concrete$ and a state $\q$, and returns datapoints. $k \in [n; 1]$ means that in the first iteration $k = n > 0$. }
  \label{alg:cegdc}
  \begin{algorithmic}[1]
  \Procedure{CEGDC}{$\concrete$, $\q$}
  \State \ \ $\mathsf{datapoints} \leftarrow [\ ]$;
  \State \ \ \textbf{for each}\ $k \in [len(\concrete);\ 1]$; \label{algo2:line1}
  \State \ \ \ \ $\q'_e \leftarrow \textsc{Estimate}(\q, \concrete, k);$ \label{algo2:line2}
  \State \ \ \ \ $\q'_a \leftarrow \textsc{Execute}(\q, \concrete, k);$  \label{algo2:line3}
  \State \ \ \ \ \textbf{if}\ $\textsc{IsAccurate}(\q'_{e}, \q'_{a})$  \label{algo2:line4}
  \State \ \ \ \ \ \ \textbf{returns}\ $\mathsf{datapoints}$ ; \label{algo2:line5}       
  \State \ \ \ \ \textbf{else}
  \State \ \ \ \ \ \ $(\act,\textsf{paras}) \leftarrow \concrete[k]$; \label{algo2:line7}
  \State \ \ \ \ \ \ $\mathsf{datapoints}[\act] \leftarrow (\q,\textsf{paras},\q'_{a})$; \label{algo2:line8}
  \State \ \ \textbf{return}\ $\mathsf{datapoints}$;
  \EndProcedure
  \end{algorithmic}
\end{algorithm}
\vspace*{-5mm}

In Algorithm~\ref{alg:cegdc}, we present the sub-procedure \textsc{CEGDC} for collecting new data points from a counterexample. \textsc{CEGDC} takes as inputs a counterexample $\concrete$ which is known to have an inaccurate profit estimation, and a blockchain state. The for loop on line~\ref{algo2:line1} is used to locate approximation errors backward from the last action to the first action and collect new data points accordingly. In a loop iteration $k$, {\name} checks if the estimated methods of the action at the index $k$ of $\concrete$ are accurate. First, {\name} computes the estimated state $\q'_e$ reached by executing $\concrete$ until reaching the action indexed $k$ (line~\ref{algo1:line2}) using the approximated transition functions. 
Second, {\name} fetches the actual state $\q'_a$ reached by executing $\concrete$ until reaching the action indexed $k$ (line~\ref{algo1:line3}) on the actual smart contracts on the blockchain. 
Then, {\name} compares the estimated and actual execution results (line~\ref{algo1:line4}). 
If the estimation is accurate, this indicates that the transition functions of the action at the index $k$ of $\concrete$ and its predecessors are accurate; so the procedure breaks the loop and returns the data points computed in the previous iterations (line~\ref{algo2:line5}). Otherwise, it indicates inaccurate transition functions of this action or/and its predecessors. Thus, we add a new data point associated with the action at the index $k$ of $\concrete$ (lines~\ref{algo2:line7} and \ref{algo2:line8}) and proceed to the next iteration of the loop to explore the action predecessors.

\subsection{{\gen}: Action Candidates Identification}

Flash loan attacks typically focus on victim contracts containing functions capable of transferring tokens,\footnote{Here, tokens refer to various forms of DeFi tokens, including stable coins, debt tokens, share tokens, liquidity tokens, asset tokens, etc.} which can be invoked by regular users. The attacker manipulates the transfer amount under specific conditions to make profit. 


We designed and built a tool {\gen} to assist users of {\name} to select action candidates likely to be involved in an attack vector. Given a set of target smart contracts, {\gen} identifies action candidates in the following steps.

\vspace{-2mm}
\subsubsection{Selecting Action Candidates from Contract Application Binary Interfaces (ABIs)}
\label{subsec:abi}
For all verified smart contracts, their ABIs are made public to facilitate users to call functions and engage with the contracts. An ABI typically comprises (public or external) function names, argument names/types, function state mutability, and return types. During the process of selecting action candidates, certain functions can be safely ignored: (1) Functions with the view or pure mutability can be excluded. (2) Functions that can only be invoked by privileged users, such as \emph{transferOwnership} and \emph{changeAdmin}, can also be disregarded since they are unlikely to be accessed by regular users\footnote{Previous works~\cite{ghaleb2023achecker, liu2022finding} have extensively researched access control vulnerabilities. We exclude them from the scope of this work.}. (3) Token permission management functions/parameters, such as the function \emph{approve} or parameter \textit{deadline}, are excluded\footnote{To simplify the search process of {\name}, these permissions are assumed to be granted maximally.}. These functions/parameters solely control whether a transaction will be reverted or not and do not affect the behaviors of contracts. 

\vspace{-2mm}
\subsubsection{Learning Special Parameters from Transaction History}
After selecting a set of action candidates from contracts' ABIs, some non-integer parameters(eg. bytes, string, address, array, enum) can still be unknown. {\gen} collects past transactions of the target contracts and extracts function level trace data from these transactions, and utilizes the trace data to learn the special parameters from previous function calls made to the contract.


\vspace{-2mm}
\subsubsection{Local Execution and Intra-dependency Analysis}
After learning special parameters, each action candidate is executed at a given block to verify its executability. An action candidate may not be executable due to various reasons: (1) the function is disabled by the owner or admin; (2) internal function calls to other contracts are disabled by the owners or admins of those contracts; (3) the function is not valid under current blockchain states. The inexecutability due to these reasons cannot be identified by static analysis and can only be determined by executions. All such inexecutable functions are filtered out.

{\gen} automatically collects storage read/write information during the execution of these functions and infers the Read-After-Write (RAW) dependencies\footnote{This RAW dependency information is also employed in {\name}'s initial data collection to expand the range of data points.} between different action candidates. An action A has a RAW dependency (or equivalently, is RAW dependent) on action B if the execution of action A reads the storage written by action B\footnote{It is important to note that this step excludes any tx.origin/msg.sender-related storage reads/writes, as such storage accesses do not alter the global state of the protocol and are therefore unlikely to impact the functional behaviors of actions.}. From the RAW dependencies, it is possible to observe that certain functions behave independently, meaning they do not have any RAW dependencies on other functions, and other functions do not have any RAW dependencies on them either. Consequently, these independent functions can be safely ignored.

After analyzing ABIs, transaction history, and local executions, {\gen} generates a list of action candidates with only their integer arguments left undetermined. These action candidates are then input into {\name} for further synthesis.

\section{Implementation}
\label{sec:impl}

\newcommand*{\runner}{\textit{Runner}}
\newcommand*{\synthesizer}{\textit{Synthesizer}}
\newcommand*{\approximators}{\textit{Approximator}}
\newcommand*{\optimizer}{\textit{Optimizer}}

\begin{figure}[h]
    \centering
    \includegraphics[width=\linewidth]{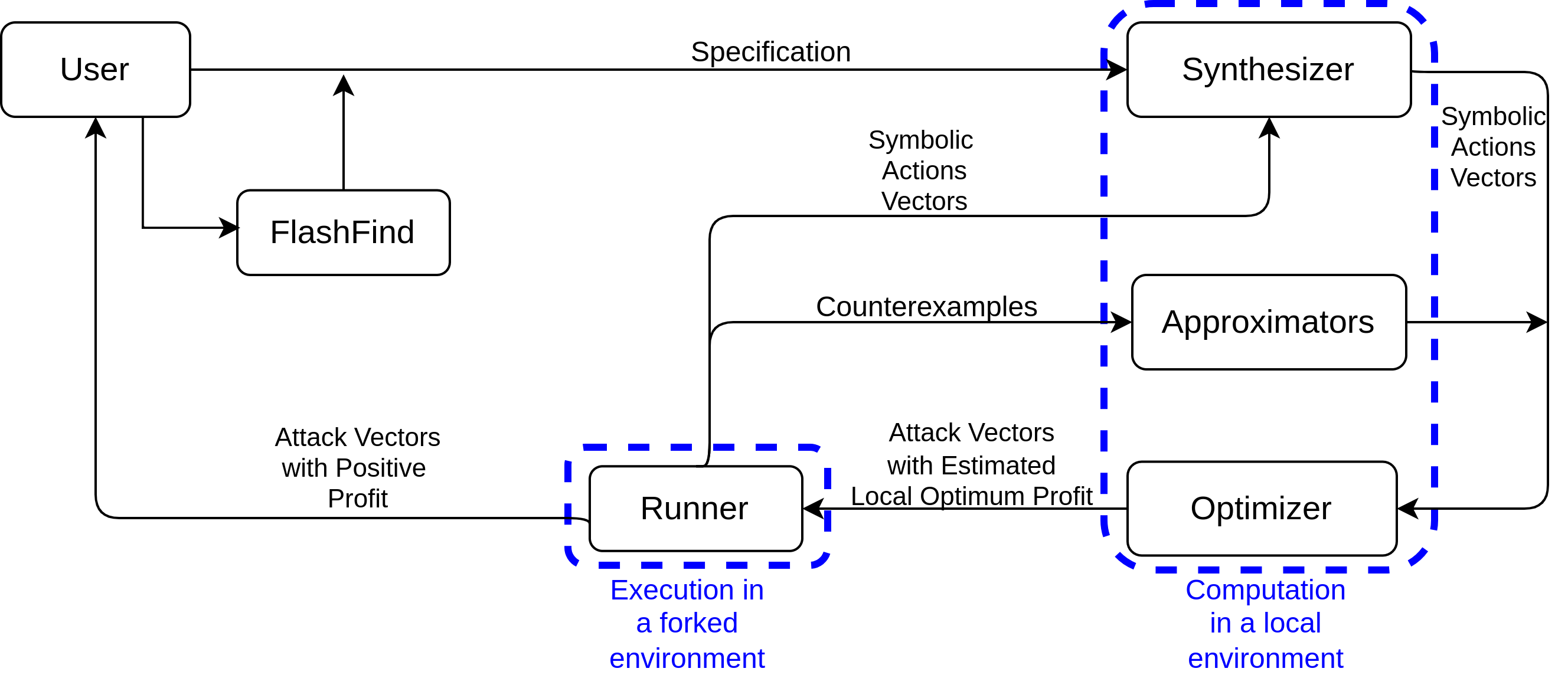}
    \vspace{-5mm}
    \caption{An Overview of {\name} Implementation}
    \vspace{-3mm}
    \label{fig:overview}
\end{figure}

{\name} is implemented in Python. 
Figure~\ref{fig:overview} shows an overview of our implementation.
\vspace{1mm}
The components {\runner}, {\synthesizer}, {\approximators}, and {\optimizer} implement the {\name} synthesis procedure presented in Algorithm~\ref{alg:enu}. An optional component {\gen} is used to automatically identify action candidates.

\noindent
\textbf{(1) {\approximators.}}
{\approximators} approximates the transition functions using data points collected by {\runner}. The approximated transition functions are then given to {\optimizer} to construct the optimization framework.  In {\approximators}, all transition functions of an action are approximated unless the transition functions 
are straightforward assignment/addition/subtraction (by simple inspection of smart contract codes), 
or the action is very common (such as Uniswap that has been widely studied~\cite{qin2021attacking,xia2021trade,xu2021sok}). 
{\name} implements two numerical methods using external libraries. {\name}-poly utilizes sklearn~\cite{scikit-learn,sklearn_api}, and {\name}-inter employs scipy~\cite{2020SciPy-NMeth}. 
\revise{The choice of polynomial and interpolation methods is motivated by several considerations. First, {\name} requires fast evaluation of approximated functions, as thousands of evaluations are performed in the optimization process. Second, when provided with an input not seen before, the approximation method needs to yield a reasonable estimation based on the nearest points. Lastly, given that a typical {\name} process  involves learning dozens of approximated formulas, the approximation process for one formula should not exceed a few seconds. Polynomial and interpolation methods are the two most popular approximation approaches that meet all of these criteria and there are off-the-shelf tools like sklearn and scipy that are easy to intergate in {\name}.}

\noindent
\textbf{(2) {\optimizer.}}
{\optimizer} automatically builds an optimization problem using the 
approximated transition functions returned by {\approximators} and the symbolic actions vectors enumerated 
by {\synthesizer} and performs global optimization on it.  The obtained attack vector that yields a positive 
profit is then executed by {\runner} to confirm the accuracy of the estimation. If the estimation is inaccurate,
the attack vector is treated as a counterexample and is used to collect new data points by {\runner} that are used by {\approximators} to refine the approximation.
We built {\optimizer} on top of an off-the-self
global optimizer \textbf{scipy.optimize.shgo}~\cite{joe2008constructing,endres2018simplicial,wales2015perspective}, which solves the simplicial 
homology global optimization algorithm to find the optimal parameters\revise{.}\footnote{Note that we were not able to use local optimizers in \textbf{scipy.optimize} library~\cite{2020SciPy-NMeth} which require an initial guess of parameters. 
Under our settings, it is not feasible to find an initial guess for every symbolic attack, as each attack behaves differently.}  In {\name}-poly, we parallelized {\optimizer} component using up to $18$ processes where {\optimizer} is run over multiple symbolic 
actions vectors in parallel.

\noindent
\textbf{(3) {\runner.}}
{\runner} executes transactions on a forked blockchain. It performs both initial and counterexample based data collection, and validates results of {\optimizer}. We implemented {\runner} on top 
of Foundry~\cite{Foundry}, a toolkit written in Rust for smart contracts development that allow to interact with EVM based blockchains.

\noindent
\textbf{(4) {\synthesizer.}}
{\synthesizer} first enumerates and prunes symbolic actions vectors using heuristics. Then, during counterexample guided 
loops, it employs priority scoring to gradually drop actions vectors based on their scores. {\synthesizer} uses iterative synthesis.
{\optimizer} can be configured with different hyperparameters to perform different strengths of parameter search. We designed $3$ sets of hyperparameters which represent different strengths of parameter search. {\synthesizer} first conducts a weakest parameter search on all enumerated symbolic actions vectors using {\optimizer}. After {\runner} validates the results, {\synthesizer} ranks symbolic actions vectors and drops the ones with low priority scores. The actions vector with high priority score will be searched with higher strengths. 
Specifically, if a symbolic actions vector yields a positive profit $\profit$ in iteration $k$, 
its priority score of iteration $k+1$ is $\profit$. If a symbolic actions vector does not yield a positive 
profit in iteration $k$, it is given a small priority score between $1$ and $10$ based on {\optimizer} results. 
An actions vector will also be dropped when its priority score does not increase between iterations. When all actions vectors are dropped, the whole synthesis procedure stops, and {\name} returns all the profitable attack vectors it found.




\vspace{1mm}
\noindent
\textbf{(5) {\gen.}} 
We also implemented {\gen} as an optional component. {\gen} uses TrueBlocks~\cite{trueblocks} and Blockchain Explorers (Etherscan~\cite{etherscan}, BscScan~\cite{bscscan}, and FtmScan~\cite{FTMScan}) to collect past transactions of the target contracts.  
{\gen} employs Phalcon~\cite{phalcon} and Foundry~\cite{Foundry} to extract function level trace data from those transactions and perform analysis on storage accesses. Our evaluation shows that {\gen} is able to identify action candidates and helps {\name} discovers alternative attack vectors (see RQ4 in Section~\ref{sec:evaluation}).

\revise{
{\name} does not require prior knowledge of a vulnerable location or contract. Given a set of DeFi lego user interface contracts, action candidates and their special parameters such as strings are given by the users or automatically extracted from transaction history using {\gen}. {\name} utilizes these action candidates to synthesize attack vectors and search for optimal numerical values. Note that these action candidates are not necessarily the ones that contain the vulnerability. Rather, they serve as user interfaces for interacting with the protocol. The vulnerability may reside in any contract invoked through nested calls originating from these action candidates. If {\gen} is not utilized, users can consult the protocol documentation to identify the appropriate user-interface contracts and functions (action candidates), as well as how to select special parameters for invoking these functions. Such information is essential for any user interacting with the protocol, and is generally available in the documentations of DeFi protocols. 
}

\setlength{\textfloatsep}{10pt}
\begin{table}[]
    \caption{Benchmark of Attacks Used in the Evaluation.}
    \centering
    \resizebox{\columnwidth}{!}{%
    \begin{tabular}{|c|c|c|c|c|c|c|}
    \hline
    \rowcolor[rgb]{ .851,  .851,  .851} 
                          & Benchmark     & \revise{\#C}$^{+}$&  \revise{LoC}$^{*}$   & Vulnerability Type       & Tx \\ \hline
    \multirow{9}{*}{ETH}  & bZx1          & \revise{6}        &  \revise{4964 }       & pump\&arbitrage       & \cite{bZx1} \\ \cline{2-6} 
                          & Harvest\_USDT & \revise{6}        &  \revise{6446 }       & manipulate oracle        & \cite{HarvestUSDC} \\ \cline{2-6} 
                          & Harvest\_USDC & \revise{6}        &  \revise{4095 }       & manipulate oracle        & \cite{HarvestUSDT}  \\ \cline{2-6} 
                          & Eminence      & \revise{2}        &  \revise{489  }       & design flaw$^{\dagger}$    & \cite{Eminence} \\ \cline{2-6} 
                          & ValueDeFi     & \revise{8}        &  \revise{7043 }       & manipulate oracle        & \cite{ValueDeFi}\\ \cline{2-6} 
                          & Cheesebank    & \revise{12}       &  \revise{1246 }       & manipulate oracle        & \cite{Cheesebank} \\ \cline{2-6} 
                          & Warp          & \revise{11}       &  \revise{13139}       & manipulate oracle        & \cite{Warp} \\ \cline{2-6} 
                          & Yearn         & \revise{5}        &  \revise{2200 }       & forced investment          & \cite{Yearn} \\ \cline{2-6} 
                          & inverseFi     & \revise{7}        &  \revise{5734 }       & manipulate oracle        & \cite{inverseFi} \\ \hline
    \multirow{6}{*}{BSC}  & bEarnFi       & \revise{3}        &  \revise{3007 }       & asset mismatch             & \cite{bEarnFi} \\ \cline{2-6} 
                          & AutoShark     & \revise{6}        &  \revise{8052 }       & design flaw$^{\dagger}$    & \cite{AutoShark} \\ \cline{2-6}
                          & ElevenFi      & \revise{7}        &  \revise{5613 }       & design flaw$^{\dagger}$    & \cite{ElevenFi} \\ \cline{2-6} 
                          & ApeRocket     & \revise{7}        &  \revise{1562 }       & design flaw$^{\dagger}$    & \cite{ApeRocket} \\ \cline{2-6} 
                          & WDoge         & \revise{2}        &  \revise{788  }       & deflationary token         & \cite{WDoge} \\ \cline{2-6} 
                          & Novo          & \revise{4}        &  \revise{7080 }       & design flaw$^{\dagger}$    & \cite{Novo}  \\ \hline
    FTM                   & OneRing       & \revise{14}       &  \revise{5386 }       & design flaw$^{\dagger}$    & \cite{OneRing} \\ \hline
    \multirow{2}{*}{DVD}  & Puppet        & \revise{2}        &  \revise{742  }       & manipulate oracle        & \cite{DVD1} \\ \cline{2-6} 
                          & PuppetV2      & \revise{1}        &  \revise{161  }       & manipulate oracle        & \cite{DVD2} \\ \hline
    \rowcolor[rgb]{ .949,  .949,  .949} 
    \multicolumn{4}{|c|}{Total Financial Loss in History}     & \multicolumn{2}{c|}{82.5 million USD} \\ \hline                        
    \end{tabular}

    }

    \flushleft
    \small
    \revise{$+$: \#C denotes number of the victim protocol's contracts invoked in exploits.}

    \revise{$*$: LoC denotes total number of lines of code in the contracts identified by \#C, excluding closed-source contracts.}

    $\dagger$: The logic designs of one or more functions in the victim contracts is flawed, with highly specific case-by-case vulnerabilities.    
    \label{tab:benchmark}
\end{table}

\section{Evaluation}
\label{sec:evaluation}
We aim to answer the following research questions:
\begin{enumerate}[label=\textbf{RQ\,\arabic*:}, ref={RQ\,\arabic*}]

\item How effective is {\name} in synthesizing flash loan attack vectors? \label{RQ1}

\item How well does the synthesis-via-approximation technique perform compared to precise baselines? \label{RQ2}

\item How much does counterexample driven approximation
    refinement improve {\name}'s results? \label{RQ3}

\item How effective is the combination of {\gen} and {\name} to synthesize attack vectors end-to-end?  \label{RQ4}
\end{enumerate}

\noindent 
\revise{
\textbf{Scope:}
{\name} focuses on flash loan attacks that generate positive profit by
sequentially invoking functions within existing DeFi contracts. Security
attacks that require exploiting other vulnerabilities such as re-entrance or
conducting social engineering are outside the scope of {\name} and our
evaluation. The goal of {\name} is to prove the existence and the
exploitability of flash loan vulnerabilities. 
Consequently, activities such as getting and repaying the flash
loan are not part of our synthesis task.
}

\noindent 
\revise{
\textbf{Benchmarks:}
We investigated historical flash loan attacks that span from
02/14/2020 to 06/16/2022 across Ethereum, Binance Smart Chain (BSC), and Fantom
(FTM) and attempted to reproduce each of them in our environment. In the end,
we reproduced $16$ attacks that are within our scope and collected them as our
benchmark attacks. 
These attacks invoked $2$-$14$ contracts of the victim protocol in the nested
invocation tree per attack, consisting of a total of $489$ to $13,139$ lines of code,
reflecting the multifaceted nature of the DeFi protocols exploited 
in real-world flash loan attacks. Also, protocols in our benchmark 
contain up to $15$ action candidates from which {\name} needs to find an attack vector. 
Altogether, the $16$ historical flash loan attacks in our benchmark have caused over $82.5$ million
US dollars in losses and include widely-known cases such as Harvest, bZx, and
Eminence. Additionally, we include $2$ fictional attacks from the Damn Vulnerable DeFi
(DVD) challenges~\cite{DVD}. 
}



\noindent \textbf{Ground Truth:} For historical flash loan attacks, we forked
the corresponding blockchain at one block prior to the attack transaction and
replayed the attacker's attack vector as the ground truth. For DVD benchmarks, we
select community solutions as ground truth. Note that in a flash loan
attack, if the same attack vector is repeated multiple times, we remove the loop and only consider the first attack vector as
the ground truth.

\newcommand{\tnote}[1]{$^{\mathrm{#1}}$}

\setlength{\textfloatsep}{8pt}
\begin{table*}[]
    \caption{Summary of {\name} Results. \textbf{AC} denotes the number of action candidates.  
    \textbf{AP} denotes the number of action candidates to approximate. 
    \textbf{GL} and \textbf{GP} denote the length and the profit of the ground truth attack vector, respectively. 
    \textbf{IDP} and \textbf{TDP} denote the initial and total number of collected of data points, respectively.
    \textbf{Time} denotes the time spent in seconds.}
    \label{tab:RQ1}
    \centering

\tabcolsep=0.07cm
\begin{tabular}{|l|llll|l|lll|lll|l|}
    \hline
    \rowcolor[rgb]{ .851,  .851,  .851}           \multicolumn{6}{|l|}{}                            & \multicolumn{3}{c|}{\textbf{{\name}-poly}}              & \multicolumn{3}{c|}{\textbf{{\name}-inter}}      & \textbf{Precise} \\ \hline
    \rowcolor[rgb]{ .851,  .851,  .851}           \textbf{Benchmark}     & \textbf{AC} & \textbf{AP} & \textbf{GL} & \textbf{GP}  & \textbf{IDP}  & \textbf{TDP} & \textbf{Profit} & \textbf{Time}   & \textbf{TDP}  & \textbf{Profit}  & \textbf{Time} & \textbf{Profit}  \\ \hline
                                                  bZx1          & 3  & 3  & 2  & 1194                         & 5192            & 5849           & 2392                   & 422             & 6373           & $2302^\dagger$      & 441             & cs       \\
    \rowcolor[rgb]{ .949,  .949,  .949}           Harvest\_USDT & 4  & 4  & 4  & 338448                       & 8000            & 9325           & $110139^\dagger$       & 670             & 10289          & $86798^\dagger$     & 7579            & cs       \\
                                                  Harvest\_USDC & 4  & 4  & 4  & 307416                       & 8000            & 8912           & $59614^\dagger$        & 677             & 10914          & $110051^\dagger$    & 8349            & cs       \\
    \rowcolor[rgb]{ .949,  .949,  .949}           Eminence      & 4  & 4  & 5  & 1674278                      & 8000            & 8780           & 1507174                & 1191            & 8104           & /                   & /               & 1606965  \\
                                                  ValueDeFi     & 6  & 6  & 6  & 8618002                      & 12000           & 19975          & $8378194^\dagger$      & 4691            & 15758          & $6428341^\dagger$   & 11089           & cx       \\
    \rowcolor[rgb]{ .949,  .949,  .949}           CheeseBank    & 8  & 3  & 8  & 3270347                      & 2679            & 2937           & $1946291^\dagger$      & 4391            & 2715           & $1101547^\dagger$   & 10942           & $2816762^\dagger$  \\
                                                  Warp          & 6  & 3  & 6  & 1693523                      & 6000            & 6000           & $2773345^\dagger$      & 1164            & 6000           & /                   & /               & $2645640^\dagger$  \\
    \rowcolor[rgb]{ .949,  .949,  .949}           bEarnFi       & 2  & 2  & 4  & 18077                        & 4000            & 4854           & 13770                  & 470             & 4652           & 12329               & 688             & 13832    \\
                                                  AutoShark     & 8  & 3  & 8  & 1381                         & 2753            & 2753           & $1372^\dagger$         & 5484            & 2753           & /                   & /               & cx       \\
    \rowcolor[rgb]{ .949,  .949,  .949}           ElevenFi      & 5  & 2  & 5  & 129741                       & 4000            & 4070           & 129658                 & 409             & 4326           & 85811               & 898             & cx       \\ 
                                                  ApeRocket     & 7  & 3  & 6  & 1345                         & 6000            & 6402           & $1333^\dagger$         & 733             & 6235           & $1037^\dagger$      & 3238            & cs       \\ 
    \rowcolor[rgb]{ .949,  .949,  .949}           Wdoge         & 5  & 1  & 5  & 78                           & 2000            & 2001           & 75                     & 272             & 2080           & 75                  & 289             & 75       \\ 
                                                  Novo          & 4  & 2  & 4  & 24857                        & 4000            & 4164           & 20210                  & 702             & 4031           & 23084               & 861             & cx       \\
    \rowcolor[rgb]{ .949,  .949,  .949}           OneRing       & 2  & 2  & 2  & 1534752                      & 4000            & 4710           & 1814882                & 585             & 4218           & 1942188             & 367             & cx       \\
                                                  Puppet        & 3  & 3  & 2  & 89000                        & 6000            & 6301           & $89000^\dagger$        & 1203            & 6452           & $87266^\dagger$     & 1238            & $89000^\dagger$    \\
    \rowcolor[rgb]{ .949,  .949,  .949}           PuppetV2      & 4  & 3  & 3  & 953100                       & 4491            & 4836           & $747799^\dagger$       & 2441            & 5061           & $362541^\dagger$    & 2835            & $647894^\dagger$   \\ \hline
\multicolumn{6}{|l|}{}  & \multicolumn{3}{c|}{\textbf{Solved}:16/18  \textbf{Avg. Time}: 1594}  & \multicolumn{3}{c|}{\textbf{Solved}:13/18 \textbf{Avg. Time}: 3754} &  \\ \hline       
\end{tabular}

\flushleft
\small
$\dagger$: {\name}'s results include at least one attack vector that differs from the ground truth.\\

\end{table*}



\noindent \textbf{Precise Baseline:} To demonstrate the effectiveness
of synthesis-via-approximation techniques, we implemented a baseline
synthesizer that works with manual summaries of smart contract actions.
Specifically, we manually inspected all benchmarks whose relevant smart contracts that are all open-source and for each benchmark we allocated more than $4$ manual analysis hours
to extract the precise mathematical summaries. The baseline synthesizer then
uses the manually extracted precise summaries to drive the synthesis.

\noindent \textbf{Environment Setup:} 
We assume that the flash loan providers are generally available, and we do not consider the borrow and the return as the part of the synthesis task. 
\revise{To facilitate {\name} experimentation, we manually annotated the prestates and poststates for each action. The details of this annotation process are described in Appendix C. Although this manual effort is required, it's worth noting that automation of this step is possible. Techniques such as dynamic taint analysis and forward symbolic execution can be employed to automatically identify which storage variables influence the change in token balances, thereby streamlining the annotation of prestates and poststates.}
The experiments are conducted on an Ubuntu $22.04$ server, with
an AMD Ryzen Threadripper 2990WX $32$-Core Processor and $128$ GB RAM. 

\noindent
\textbf{Experiment Overview.}
\revise{To answer \ref{RQ1}, we apply {\name} to the $18$ benchmarks with the same set of candidate actions in ground truths. For each candidate action, the prestates and poststates are annotated for {\name} to
drive the approximated formula for this action.} We set a timeout of $3$ hours for
{\name}-poly and $4$ hours for {\name}-inter. {\name} does not know a priori whether a benchmark has an attack vector with a positive profit, and it does not set any bounds on the profit. It tries iteratively to synthesize an attack vector with a maximum profit. {\name}'s refinement loop is guided by intermediate results and {\name} stops when it cannot improve the profit or the above timeouts are reached. To answer \ref{RQ2}, we replace {\approximators} component of {\name} with manually extracted precise mathematical summaries, and conduct the same experiment with $4$ hours timeout. To answer \ref{RQ3}, we evaluate {\name} with different initial data points and with CEGDC enabled/disabled. \revise{To answer \ref{RQ4}, we first use {\gen} to identify candidate actions from given contract addresses which the hacker used in history, manually annotate them as in \ref{RQ1}, and then apply {\name} with this new set of candidate actions to synthesize attack vectors under the same setting as in \ref{RQ1}.} The results for \ref{RQ1}+\ref{RQ2}, \ref{RQ3}, \ref{RQ4} are summarized in Table~\ref{tab:RQ1}, Table~\ref{tab:RQ3Summary}, and Table~\ref{tab:RQ4}, respectively.

\setlength{\textfloatsep}{10pt}
\begin{table*}[]
    \caption{Summary of {\name} Results under Different Settings (see Appendix F for complete results). 
        \textbf{n+x}: \textbf{n} denotes the settings of initial number of data points and \textbf{+x} denotes whether {\name} uses counterexample driven loops. }
    \centering
    \setlength{\tabcolsep}{2pt}
\tabcolsep=0.07cm
\begin{tabular}{|l|llllllll|llllllll|}
    \hline
\rowcolor[rgb]{ .851,  .851,  .851}                   & \multicolumn{8}{c|}{\textbf{{\name}-poly} }     & \multicolumn{8}{c|}{\textbf{{\name}-inter}} \\ \hline
\rowcolor[rgb]{ .851,  .851,  .851}                   & \textbf{200}  & \textbf{200+x} & \textbf{500} & \textbf{500+x} & \textbf{1000} & \textbf{1000+x} & \textbf{2000} & \textbf{2000+x}  &  \textbf{200}  & \textbf{200+x} & \textbf{500}  & \textbf{500+x} & \textbf{1000} & \textbf{1000+x} & \textbf{2000} & \textbf{2000+x}  \\ \hline
                                                      Avg. Time (s)     & 632  & 893   & 1120 & 1747  & 842  & 1397   & 982  & 1594  & 2601 & 3509  & 3180 & 3917  & 3022 & 3845   & 3200 & 3754    \\
\rowcolor[rgb]{ .949,  .949,  .949}                   Avg. Data Points  & 584  & 1042  & 1432 & 2376  & 2795 & 3571   & 5445 & 6367  & 584  & 1338  & 1432 & 2450  & 2795 & 3656   & 5445 & 6248   \\
                                                      Avg. Norm. Profit & 0.793 & 0.829  & 0.846 & 0.922  & 0.762 & 0.786  & 0.717 & 0.945 & 0.539 & 0.555  & 0.630 & 0.634  & 0.535 & 0.580   & 0.594 & 0.641    \\ 
\rowcolor[rgb]{ .949,  .949,  .949}                   Benchmarks Solved  & 15   & 15    & 15   & 16    & 15   & 15     & 15   & 16    & 13   & 13    & 14   & 14    & 13   & 13     & 13   & 13  \\ \hline
\end{tabular}
\label{tab:RQ3Summary}
\vspace{-3mm}
\end{table*}

\noindent
\textbf{RQ1: Effectiveness of {\name}.}
Table~\ref{tab:RQ1} summarizes the results of the experiment. The first five columns of Table~\ref{tab:RQ1} list benchmark information including the number of actions to be approximated and the length of the ground truths. The
four columns under {\name}-poly list data concerning the synthesis using polynomial approximations. The four columns under {\name}-inter
list data concerning the synthesis using interpolation based approximation.

Our results in Table~\ref{tab:RQ1} show that {\name} can effectively synthesize
flash loan attack vectors. {\name}-poly
(resp., {\name}-inter) synthesizes profitable attack vectors for $16$ (resp.,
$13$) benchmarks with an average normalized profit (w.r.t. the ground
truth profit) of $0.945$ (resp., $0.641$). In particular, for three benchmarks
(\emph{ApeRocket}, \emph{ElevenFi}, and \emph{AutoShark}) the profits found by
{\name}-poly are within $99\%$ of the profits in the original attacks vectors.
Surprisingly in another three benchmarks (\emph{bZx1}, \emph{Warp}, and \emph{OneRing})
the profits found by {\name} are bigger than the profits in the original
attacks vectors. For instance, in the \emph{Warp} benchmark the profit is roughly double the ground truth profit (see Appendix B for Warp case study). 
On average, {\name}-poly is $\times2$ faster than {\name}-inter, because we used parallelism in {\name}-poly which is not possible for {\name}-inter. 

For $10$ benchmarks, {\name} successfully discovers new profitable symbolic actions vectors that are different from the ground truths. These vectors either exploit the same vulnerability but in a different order of actions, or represent arbitrage opportunities that were not exploited by the original attackers. For the remaining $6$ benchmarks, {\name} discovers exactly the same symbolic actions vectors as the ground truths but with different parameters. Note that {\name} is not able to solve \emph{Yearn} and \emph{InverseFi} which are not shown in Table~\ref{tab:RQ1}. These two benchmarks 
put high requirements on the precision of the approximation and small miss-approximation errors caused {\name} to miss finding attack vectors and accurate parameters.  

\revise{
    To evaluate the efficacy of the pruning heuristics introduced in Section ~\ref{sec:isfeasible}, we conduct experiments comparing the search space sizes when using {\name} with and without the application of some of the heuristics. Our results indicate that Heuristic 1 leads to an average reduction of $57\%$ in the search space size. Subsequently, Heuristic 2 further reduces the remaining search space by an additional $34\%$, and Heuristic 3 contributes an additional reduction of $65\%$ to the remaining search space.
}

\revise{
To compare {\name} with existing static analyzers, we manually select contracts
containing vulnerabilities in the benchmarks and apply the popular smart contracts static analyzer
Slither~\cite{feist2019slither} to them. In the experiments, we identify contracts that contain the root cause of the vulnerabilities as the target contracts for Slither to analyze.
Note that in practice, identifying target contracts for Slither is much harder
than that for {\gen}. For {\gen}, the target contracts are simply
user-interface contracts. In contrast, identifying the contract with the actual
vulnerability, such as a contract invoked in a deeply nested call chain, 
can be tedious. 
Slither fails to detect vulnerabilities for all $18$ benchmarks, among them
Slither fails to parse $2$ benchmarks (Novo and Yearn). The possible reasons
include: (i) Slither's inability to reason across multi-contract interactions,
common in flash loan attacks; and (ii) its lack of context awareness, such as
not detecting Uniswap~\cite{Uniswap} when used as an oracle.
}

\noindent
\textbf{RQ2: Comparison with Precise Baseline.}
The last column of Table~\ref{tab:RQ1} lists data when the approximation component of {\name} is replaced with precise mathematical summaries
for actions. Note that $4$ benchmarks are partially closed-source (\textbf{cs}), and $5$ benchmarks are too complicated (\textbf{cx}), thus we are not able to extract mathematical precise summaries for them. For others, we list the profit generated using the manually extracted mathematical expressions in the synthesizer and optimizer. 

\justify
Our results in Table~\ref{tab:RQ1} show that the \revise{synthesis-via-approxima- tion}
approach performs well compared to precise baselines. For the $9$ cases that
the precise baseline failed due to either close source (\textbf{cs}) or
complicated contract logics (\textbf{cx}), {\name} found attack vectors that
generate positive profits. On average, for the $7$ cases that
the precise baseline succeeds, the best profit from {\name} is $0.97$
of the profit returned by the precise baseline. In particular, for \emph{Warp} and \emph{PuppetV2}
{\name} synthesizes an attack vector with a profit higher than that
obtained by the precise approach. This is because the approximations 
used in {\name} are simpler than the mathematical summaries used in precise baseline. 
This enables the optimizer to converge faster and find better parameter values within the fixed time budget.

\noindent
\textbf{RQ3: Counterexample Driven Approximation Refinement.}
Table~\ref{tab:RQ3Summary} summarizes the evaluation of {\name} under different settings. In
particular, we evaluated {\name} with $200$, $500$, $1000$, and $2000$ initial
data points threshold per action to be approximated without and with
counterexample loop. 
The \textbf{Avg.} rows in Table~\ref{tab:RQ3Summary} are calculated based on the $16$ benchmarks
excluding \emph{Yearn} and \emph{InverseFi}. The \textbf{Avg. Norm. 
Profit} is calculated as the average of normalized profits, i.e., profit /
ground truth profit. 

For {\name}-poly, the results in Table~\ref{tab:RQ3Summary} show that only
with counterexample loop we are able to solve the $16$ benchmarks (Columns
\textbf{500+x} and \textbf{2000+x}). Also, the maximum average of normalized
profits is achieved with counterexample loop (Column \textbf{2000+x}) which
improved from $0.717$ (Column \textbf{2000}) without counterexample loop to
$0.945$ with counterexample loop. For {\name}-inter, the maximum
average of normalized profits is also achieved with counterexample loop (Column
\textbf{2000+x}) which improved from $0.594$ (Column \textbf{2000}) without
counterexample loop to $0.641$ with counterexample loop. 



\setlength{\textfloatsep}{8pt}
\begin{table*}[]
    \caption{Summary of Evaluation Results of Combining {\name} with {\gen}. \textbf{AC} is the number of action candidates.  
    \textbf{AP} is the number of action candidates to approximate. 
    \textbf{GL} and \textbf{GP} are the length and the profit of the ground truth attack vector, respectively. 
    \textbf{IDP} and \textbf{TDP} are the initial and total number of collected of data points, respectively.
    \textbf{Time} is measured in seconds.}
    \label{tab:RQ4}
    \centering
\tabcolsep=0.07cm
\begin{tabular}{|l|ll|llllll|llllll|}
    \hline
    \rowcolor[rgb]{ .851,  .851,  .851}
    \multicolumn{3}{|l|}{}                            & \multicolumn{6}{c|}{ \textbf{{\gen} + {\name}-poly}}        & \multicolumn{6}{c|}{ \textbf{{\name}-poly} }       \\ \hline
    \rowcolor[rgb]{ .851,  .851,  .851}
    \textbf{Benchmark}     & \textbf{GL} & \textbf{GP}  & \textbf{AC} & \textbf{AP} & \textbf{IDP}  & \textbf{TDP} & \textbf{Profit} & \textbf{Time} &  \textbf{AC} & \textbf{AP}  & \textbf{IDP}  & \textbf{TDP}  & \textbf{Profit}  & \textbf{Time}  \\  \hline

bZx1          & 2  & 1194                   & 3  & 3        & 5192            & 5849           & 2392                    & 422           & 3  & 3       & 5192            & 5849           & 2392             & 422         \\

\rowcolor[rgb]{ .949,  .949,  .949}
Harvest\_USDT & 4  & 338448                 & 15  & 15        & 30000           & 34052          & $85593^\ddagger$        & 5514           & 4  & 4       & 8000            & 9325           & $110139^\dagger$           & 670         \\
Harvest\_USDC & 4  & 307416                 & 15  & 15        & 30000           & 51726          & $33645^\ddagger$       & 3630           & 4  & 4       & 8000            & 8912           & $59614^\dagger$           & 677         \\

\rowcolor[rgb]{ .949,  .949,  .949}
Eminence      & 4  & 1674278                & 4  & 4        & 8000            & 8780           & 1507174                 & 1191           & 4  & 4       & 8000            & 8780           & 1507174         & 1191         \\
ValueDeFi     & 6  & 8618002                & 6  & 6        & 12000           & 19975          & $8378194^\dagger$       & 4691           & 6  & 6       & 12000           & 19975          & $8378194^\dagger$        & 4691         \\

\rowcolor[rgb]{ .949,  .949,  .949}
Warp          & 6  & 1693523                & 8  & 5        & 7772            & 7772           & $2776351^\ddagger$       & 3129           & 6  & 3       & 6000            & 6000           & $2776351^\dagger$         & 1164         \\
bEarnFi       & 4  & 18077                  & 2  & 2        & 4000            & 4854           & 13770                   & 470            & 2  & 2       & 4000            & 4854           & 13770           & 470         \\

\rowcolor[rgb]{ .949,  .949,  .949}
ApeRocket     &   6 & 1345                  & 11 & 5         & 10000       & 10706               & $1179^\ddagger$       & 3064           & 7  & 3       & 6000            & 6402           & $1333^\dagger$            & 733         \\ 
Wdoge         &   5 & 78                    & 7  & 2         & 4000        & 4107               & $75^\ddagger$          & 769            & 5  & 1       & 2000            & 2001           & 75              & 272         \\ 

\rowcolor[rgb]{ .949,  .949,  .949}
Novo          &   4 & 24857                 & 6  & 2         & 4000        & 4172               &  15183                 & 791            & 4  & 2       & 4000            & 4164           & 20210           & 702         \\
OneRing       &   2 & 1534752               & 8  & 8         & 16000       & 16614              &  $1814877^\ddagger$    & 1104           & 2  & 2       & 4000            & 4710           & 1814882         & 585         \\
\hline
\end{tabular}

\flushleft
\small

$\dagger$: {\name}'s results include at least one attack vector that differs from the ground truth.\\
$\ddagger$: {\name}'s results include at least one attack vector that contains an action not present in the ground truth.
\vspace{-3mm}
\end{table*} 

\noindent
\textbf{RQ4: Effectiveness of {\gen}.}
In this experiment, we evaluate the combination of {\gen} and {\name} on the $14$ benchmarks that {\name} was able to synthesize a profitable attack vector in Table~\ref{tab:RQ1} excluding the two fictional DVD benchmarks\revise{.}\footnote{DVD benchmarks do not have historical transactions that {\gen} can use.}. In particular, only contract addresses are provided to {\gen} and {\gen} identifies candidate actions for {\name} to synthesize attack vectors with the $2000$ initial data points threshold per action configuration. Table \ref{tab:RQ4} presents the results. 

{\gen} successfully identifies a reasonable number of action candidates for $11$ out of the $14$ benchmarks from given contract addresses. Among them, {\gen} identifies additional candidate actions for $7$ benchmarks. For instance, {\gen} identifies $6$ additional candidate actions for \emph{OneRing}. The remaining $3$ benchmarks contains action candidates whose arguments are non-primitive types, and {\gen} identifies an excessive and impractical number of choices from transaction history\revise{.}\footnote{In such cases, we believe experienced security analysts could manually identify special parameters and further reduce the number of parameter choices.}

Even with the extra candidate actions {\name} was able to synthesize profitable attack vectors for all $11$ benchmarks.   
Surprisingly in $6$ benchmarks, {\name} finds attack vectors that contains new action candidates from {\gen} that are not in the ground truth. There are two possibilities: First, the new action candidates identified by {\gen} are functionally similar to one action in ground truths (e.g, \codeword{withdraw} and \codeword{withdrawSafe} for \emph{OneRing}). Replacing old actions with new ones gives new attack vectors. Second, the new action candidates represent another way of draining assets which the attacker failed to identify. For example, in the \emph{Warp} benchmark, the attacker only invoked \emph{borrowSC(USDC, v)} and \emph{borrowSC(DAI, v)} to drain USDC and DAI~\cite{Warp, Rekt}, however, {\gen} identifies \emph{borrowSC(USDT, v)} as another candidate action, which could have been used to drain USDT as well in the same transaction. 


\noindent
\textbf{Impact.} 
\revise{One author of this paper collaborated with Quantstamp for applying {\name} for 3 months. We discovered two zero-day flash loan vulnerabilities in two protocols under audit.}

\noindent
\textbf{Threats to Validity:}
The \textit{internal} threat to validity mainly lies in human mistakes in the study. Specifically, we may understand results of {\name} incorrectly, or make mistakes in the implementation of {\name}. All authors have extensive smart contract security analysis experience and software engineering expertise in general. To further reduce this threat, we manually check the balance changes for the best results given by {\name} in each benchmark. We verify that with the help of on-chain exchanges, these attack vectors can generate a post-balance strictly larger than initial capital (see Appendix E). 
The \textit{external} threat to validity mainly lies in the subjects used in our study. The flash loan attacks we study might not be representative. \revise{We mitigate this risk by using diverse and reputable data sources, including academic papers~\cite{qin2021attacking,cao2021flashot} and an industrial database~\cite{SlowMistStats}.}

\noindent
\revise{
\textbf{Limitations:}
Like most synthesis tools, {\name} faces scalability challenges. The search space grows exponentially with the number of actions and attack vector length. A practical approach is to assess protocols on a module by module basis. By focusing only on inter-dependent actions within, we can maintain both the number of actions and the attack vector length at manageable levels, thereby mitigating the scaling issue.
}

\vspace{-2mm}
\section{Related Work}
\label{sec:related}

\noindent \textbf{Parametric optimization:} For some flash loan attack cases,
researchers~\cite{cao2021flashot, qin2021attacking} manually extracted math formulas
of function candidates, manually defined related parameter constraints, and 
used an off-the-shelf optimizer to search for parameters which yield the best profit. 
However, this technique requires
significant manual efforts and expert knowledge of the underlying DeFi
protocols. Consequently, it becomes impractical for checking a large number of potential attack vectors. 
Note that our benchmark set contains significantly more flash loan attacks than 
prior work~\cite{qin2021attacking,cao2021flashot}, i.e., $18$ versus $2$ in~\cite{qin2021attacking} and $9$ in~\cite{cao2021flashot}. 

\noindent \textbf{Static Analysis:} 
Slither~\cite{feist2019slither}, Securify~\cite{tsankov2018securify},
Zeus~\cite{kalra2018zeus}, Park~\cite{zheng2022park} and
SmartCheck~\cite{tikhomirov2018smartcheck} apply static analysis techniques to
verify smart contracts. 
There are also several works that use
symbolic execution~\cite{king1976symbolic} to explore the program states of a
smart contract, looking for an execution path that violates a user-defined
invariant, e.g., Mythril~\cite{mythril}, Oyente~\cite{luu2016making},
FairCon~\cite{liu2020towards}, ETHBMC~\cite{frank2020ethbmc},
SmartCopy~\cite{feng2019precise}, and Manticore~\cite{mossberg2019manticore}.
These techniques tend to operate with one contract at a time and therefore cannot handle flash loan attacks that involve multiple contracts.
These techniques also may suffer from the complicated logics of the DeFi contracts, and cause
path explosion.
{\name} uses its novel synthesis-via-approximation techniques to avoid these issues.

\noindent \textbf{Fuzzing:}
ContractFuzzer~\cite{jiang2018contractfuzzer}, sFuzz~\cite{nguyen2020sfuzz},
ContraMaster~\cite{wang2020oracle}, SMARTIAN~\cite{choi2021smartian} and ItyFuzz~\cite{shou2023ityfuzz}
introduce novel fuzzing techniques to discover vulnerabilities in smart
contracts. However, these techniques either only work on one contract or focus
on specific vulnerabilities like re-entrancy. Moreover, flash loan attack vectors can contain 
up to $8$ actions and $7$ integer parameters, which is unlikely to be found by random fuzzing.

\noindent \textbf{Flash Loan Attacks:} 
Prior works~\cite{qin2021attacking,cao2021flashot,werapun2022flash} study specific flash loan attacks and 
manually analyze and optimize the attack vectors. Some other tools~\cite{xia2023detecting, ramezany2023midnight, 
wang2022defiscanner} 
are designed to monitor flash loan attacks after they happened. Other researchers investigate other usage of 
flash loans including arbitrage~\cite{wang2021towards} or wash trading~\cite{,gan2022understanding}. To the best
of our knowledge, {\name} is the first tool that can detect flash loan attacks before they happen and 
show successes in real-world DeFi protocols under auditing. 
\section{Conclusion}
\label{sec:conclusion}

We have proposed an automated synthesis framework based on numerical
approximation to generate the flash loan attack vectors on DeFi protocols.
Our results of {\name} show that the proposed framework is practical and
{\name} can automatically synthesize attack vectors for real world attacks. Our
results also highlight the effectiveness of the synthesis-via-approximation
approach. The approach helps {\name} to overcome the challenges posed by
complicated functions in DeFi protocols and our results demonstrate that using
approximations of these functions is sufficient to drive the synthesis process. 
\name{} has been adopted by a top smart contract auditing company to detect
flash loan vulnerabilities. 
The paper also points out a new promising direction for solving trace synthesis
problems when facing complicated functions.


\section{Data Availability}
The benchmarks, source code and experimental data of our artifacts are publicly accessible on \cite{chen_2024_github} and have been archived on \cite{chen_2024_Zenodo}. 


\bibliographystyle{ACM-Reference-Format}
\bibliography{main}


\begin{thebibliography}{86}


\ifx \showCODEN    \undefined \def \showCODEN     #1{\unskip}     \fi
\ifx \showDOI      \undefined \def \showDOI       #1{#1}\fi
\ifx \showISBNx    \undefined \def \showISBNx     #1{\unskip}     \fi
\ifx \showISBNxiii \undefined \def \showISBNxiii  #1{\unskip}     \fi
\ifx \showISSN     \undefined \def \showISSN      #1{\unskip}     \fi
\ifx \showLCCN     \undefined \def \showLCCN      #1{\unskip}     \fi
\ifx \shownote     \undefined \def \shownote      #1{#1}          \fi
\ifx \showarticletitle \undefined \def \showarticletitle #1{#1}   \fi
\ifx \showURL      \undefined \def \showURL       {\relax}        \fi
\providecommand\bibfield[2]{#2}
\providecommand\bibinfo[2]{#2}
\providecommand\natexlab[1]{#1}
\providecommand\showeprint[2][]{arXiv:#2}

\bibitem[1inch Network(2023)]%
        {1inch}
\bibfield{author}{\bibinfo{person}{The 1inch Network}.} \bibinfo{year}{2023}\natexlab{}.
\newblock \bibinfo{title}{The 1inch SWAP}.
\newblock \bibinfo{howpublished}{\url{https://app.1inch.io/}}.
\newblock
\newblock
\shownote{Accessed: 2023-03-25}.


\bibitem[BlockSec(2023)]%
        {phalcon}
\bibfield{author}{\bibinfo{person}{BlockSec}.} \bibinfo{year}{2023}\natexlab{}.
\newblock \bibinfo{title}{Phalcon: Powerful Transaction Explorer Designed for DeFi community}.
\newblock
\newblock
\urldef\tempurl%
\url{https://phalcon.xyz/}
\showURL{%
\tempurl}
\newblock
\shownote{Accessed: 2023-03-27}.


\bibitem[Brent et~al\mbox{.}(2020)]%
        {brent2020ethainter}
\bibfield{author}{\bibinfo{person}{Lexi Brent}, \bibinfo{person}{Neville Grech}, \bibinfo{person}{Sifis Lagouvardos}, \bibinfo{person}{Bernhard Scholz}, {and} \bibinfo{person}{Yannis Smaragdakis}.} \bibinfo{year}{2020}\natexlab{}.
\newblock \showarticletitle{Ethainter: a smart contract security analyzer for composite vulnerabilities}. In \bibinfo{booktitle}{\emph{Proceedings of the 41st ACM SIGPLAN Conference on Programming Language Design and Implementation}}. \bibinfo{pages}{454--469}.
\newblock


\bibitem[BscScan(2021a)]%
        {ApeRocket}
\bibfield{author}{\bibinfo{person}{BscScan}.} \bibinfo{year}{2021}\natexlab{a}.
\newblock \bibinfo{title}{ApeRocket Attack Transaction}.
\newblock \bibinfo{howpublished}{\url{https://bscscan.com/tx/0x701a308fba23f9b328d2cdb6c7b245f6c3063a510e0d5bc21d2477c9084f93e0}}.
\newblock


\bibitem[BscScan(2021b)]%
        {AutoShark}
\bibfield{author}{\bibinfo{person}{BscScan}.} \bibinfo{year}{2021}\natexlab{b}.
\newblock \bibinfo{title}{AutoShark Attack Transaction}.
\newblock \bibinfo{howpublished}{\url{https://bscscan.com/tx/0xfbe65ad3eed6b28d59bf6043debf1166d3420d214020ef54f12d2e0583a66f13}}.
\newblock


\bibitem[BscScan(2021c)]%
        {bEarnFi}
\bibfield{author}{\bibinfo{person}{BscScan}.} \bibinfo{year}{2021}\natexlab{c}.
\newblock \bibinfo{title}{bEarnFi Attack Transaction}.
\newblock \bibinfo{howpublished}{\url{https://bscscan.com/tx/0x603b2bbe2a7d0877b22531735ff686a7caad866f6c0435c37b7b49e4bfd9a36c}}.
\newblock


\bibitem[BscScan(2021d)]%
        {ElevenFi}
\bibfield{author}{\bibinfo{person}{BscScan}.} \bibinfo{year}{2021}\natexlab{d}.
\newblock \bibinfo{title}{ElevenFi Attack Transaction}.
\newblock \bibinfo{howpublished}{\url{https://bscscan.com/tx/0x16c87d9c4eb3bc6c4e5fbba789f72e8bbfc81b3403089294a81f31b91088fc2f}}.
\newblock


\bibitem[BscScan(2022a)]%
        {Novo}
\bibfield{author}{\bibinfo{person}{BscScan}.} \bibinfo{year}{2022}\natexlab{a}.
\newblock \bibinfo{title}{Novo Attack Transaction}.
\newblock \bibinfo{howpublished}{\url{https://bscscan.com/tx/0xc346adf14e5082e6df5aeae650f3d7f606d7e08247c2b856510766b4dfcdc57f}}.
\newblock


\bibitem[BscScan(2022b)]%
        {WDoge}
\bibfield{author}{\bibinfo{person}{BscScan}.} \bibinfo{year}{2022}\natexlab{b}.
\newblock \bibinfo{title}{WDoge Attack Transaction}.
\newblock \bibinfo{howpublished}{\url{https://bscscan.com/tx/0x4f2005e3815c15d1a9abd8588dd1464769a00414a6b7adcbfd75a5331d378e1d}}.
\newblock


\bibitem[BscScan(2023)]%
        {bscscan}
\bibfield{author}{\bibinfo{person}{BscScan}.} \bibinfo{year}{2023}\natexlab{}.
\newblock \bibinfo{title}{The BNB Smart Chain Explorer}.
\newblock
\newblock
\urldef\tempurl%
\url{https://bscscan.com/}
\showURL{%
\tempurl}


\bibitem[Buitinck et~al\mbox{.}(2013)]%
        {sklearn_api}
\bibfield{author}{\bibinfo{person}{Lars Buitinck}, \bibinfo{person}{Gilles Louppe}, \bibinfo{person}{Mathieu Blondel}, \bibinfo{person}{Fabian Pedregosa}, \bibinfo{person}{Andreas Mueller}, \bibinfo{person}{Olivier Grisel}, \bibinfo{person}{Vlad Niculae}, \bibinfo{person}{Peter Prettenhofer}, \bibinfo{person}{Alexandre Gramfort}, \bibinfo{person}{Jaques Grobler}, \bibinfo{person}{Robert Layton}, \bibinfo{person}{Jake VanderPlas}, \bibinfo{person}{Arnaud Joly}, \bibinfo{person}{Brian Holt}, {and} \bibinfo{person}{Ga{\"{e}}l Varoquaux}.} \bibinfo{year}{2013}\natexlab{}.
\newblock \showarticletitle{{API} design for machine learning software: experiences from the scikit-learn project}. In \bibinfo{booktitle}{\emph{ECML PKDD Workshop: Languages for Data Mining and Machine Learning}}. \bibinfo{pages}{108--122}.
\newblock


\bibitem[Buterin(2014)]%
        {buterin2014ethereum}
\bibfield{author}{\bibinfo{person}{Vitalik Buterin}.} \bibinfo{year}{2014}\natexlab{}.
\newblock \bibinfo{title}{Ethereum: A next-generation smart contract and decentralized application platform}.
\newblock
\newblock
\newblock
\shownote{\url{https://ethereum.org/en/whitepaper/}}.


\bibitem[Cao et~al\mbox{.}(2021)]%
        {cao2021flashot}
\bibfield{author}{\bibinfo{person}{Yixin Cao}, \bibinfo{person}{Chuanwei Zou}, {and} \bibinfo{person}{Xianfeng Cheng}.} \bibinfo{year}{2021}\natexlab{}.
\newblock \showarticletitle{Flashot: a snapshot of flash loan attack on DeFi ecosystem}.
\newblock \bibinfo{journal}{\emph{arXiv preprint arXiv:2102.00626}} (\bibinfo{year}{2021}).
\newblock


\bibitem[Chen et~al\mbox{.}(2024a)]%
        {chen_2024_github}
\bibfield{author}{\bibinfo{person}{Zhiyang Chen}, \bibinfo{person}{Sidi~Mohamed Beillahi}, {and} \bibinfo{person}{Fan Long}.} \bibinfo{year}{2024}\natexlab{a}.
\newblock \bibinfo{booktitle}{\emph{{GitHub Artifact for FlashSyn: Flash Loan Attack Synthesis via Counter Example Driven Approximation}}}.
\newblock
\urldef\tempurl%
\url{https://github.com/FlashSyn-Artifact/FlashSyn-Artifact-ICSE24}
\showURL{%
\tempurl}


\bibitem[Chen et~al\mbox{.}(2024b)]%
        {chen_2024_Zenodo}
\bibfield{author}{\bibinfo{person}{Zhiyang Chen}, \bibinfo{person}{Sidi~Mohamed Beillahi}, {and} \bibinfo{person}{Fan Long}.} \bibinfo{year}{2024}\natexlab{b}.
\newblock \bibinfo{booktitle}{\emph{{Zenodo Artifact for FlashSyn: Flash Loan Attack Synthesis via Counter Example Driven Approximation}}}.
\newblock
\urldef\tempurl%
\url{https://zenodo.org/records/10458602}
\showURL{%
\tempurl}


\bibitem[Choi et~al\mbox{.}(2021)]%
        {choi2021smartian}
\bibfield{author}{\bibinfo{person}{Jaeseung Choi}, \bibinfo{person}{Doyeon Kim}, \bibinfo{person}{Soomin Kim}, \bibinfo{person}{Gustavo Grieco}, \bibinfo{person}{Alex Groce}, {and} \bibinfo{person}{Sang~Kil Cha}.} \bibinfo{year}{2021}\natexlab{}.
\newblock \showarticletitle{SMARTIAN: Enhancing smart contract fuzzing with static and dynamic data-flow analyses}. In \bibinfo{booktitle}{\emph{2021 36th IEEE/ACM International Conference on Automated Software Engineering (ASE)}}. IEEE, \bibinfo{pages}{227--239}.
\newblock


\bibitem[Clarke et~al\mbox{.}(2000)]%
        {clarke2000counterexample}
\bibfield{author}{\bibinfo{person}{Edmund Clarke}, \bibinfo{person}{Orna Grumberg}, \bibinfo{person}{Somesh Jha}, \bibinfo{person}{Yuan Lu}, {and} \bibinfo{person}{Helmut Veith}.} \bibinfo{year}{2000}\natexlab{}.
\newblock \showarticletitle{Counterexample-guided abstraction refinement}. In \bibinfo{booktitle}{\emph{International Conference on Computer Aided Verification}}. Springer, \bibinfo{pages}{154--169}.
\newblock


\bibitem[ConsenSys(2022)]%
        {mythril}
\bibfield{author}{\bibinfo{person}{ConsenSys}.} \bibinfo{year}{2022}\natexlab{}.
\newblock \bibinfo{title}{Mythril}.
\newblock \bibinfo{howpublished}{\url{https://github.com/ConsenSys/mythril}}.
\newblock
\newblock
\shownote{Accessed: 2022-06-06}.


\bibitem[{Damn Vulnerable DeFi}(2023a)]%
        {DVD}
\bibfield{author}{\bibinfo{person}{{Damn Vulnerable DeFi}}.} \bibinfo{year}{2023}\natexlab{a}.
\newblock \bibinfo{howpublished}{\url{https://www.damnvulnerabledefi.xyz/}}.
\newblock
\newblock
\shownote{Accessed: 2023-01-31}.


\bibitem[{Damn Vulnerable DeFi}(2023b)]%
        {DVD1}
\bibfield{author}{\bibinfo{person}{{Damn Vulnerable DeFi}}.} \bibinfo{year}{2023}\natexlab{b}.
\newblock \bibinfo{title}{Challenge \#8 - Puppet}.
\newblock \bibinfo{howpublished}{\url{https://www.damnvulnerabledefi.xyz/challenges/8.html}}.
\newblock
\newblock
\shownote{Accessed: 2023-01-31}.


\bibitem[{Damn Vulnerable DeFi}(2023c)]%
        {DVD2}
\bibfield{author}{\bibinfo{person}{{Damn Vulnerable DeFi}}.} \bibinfo{year}{2023}\natexlab{c}.
\newblock \bibinfo{title}{Challenge \#9 - Puppet v2}.
\newblock \bibinfo{howpublished}{\url{https://www.damnvulnerabledefi.xyz/challenges/9.html}}.
\newblock
\newblock
\shownote{Accessed: 2023-01-31}.


\bibitem[DefiLlama(2023)]%
        {DeFillama}
\bibfield{author}{\bibinfo{person}{DefiLlama}.} \bibinfo{year}{2023}\natexlab{}.
\newblock \bibinfo{title}{DefiLlama}.
\newblock \bibinfo{howpublished}{\url{https://defillama.com/}}.
\newblock
\newblock
\shownote{Accessed: 2023-04-01}.


\bibitem[Egorov(2019)]%
        {egorov2019stableswap}
\bibfield{author}{\bibinfo{person}{Michael Egorov}.} \bibinfo{year}{2019}\natexlab{}.
\newblock \showarticletitle{StableSwap-efficient mechanism for Stablecoin liquidity}.
\newblock \bibinfo{journal}{\emph{Retrieved Feb}}  \bibinfo{volume}{24} (\bibinfo{year}{2019}), \bibinfo{pages}{2021}.
\newblock


\bibitem[Endres et~al\mbox{.}(2018)]%
        {endres2018simplicial}
\bibfield{author}{\bibinfo{person}{Stefan~C Endres}, \bibinfo{person}{Carl Sandrock}, {and} \bibinfo{person}{Walter~W Focke}.} \bibinfo{year}{2018}\natexlab{}.
\newblock \showarticletitle{A simplicial homology algorithm for Lipschitz optimisation}.
\newblock \bibinfo{journal}{\emph{Journal of Global Optimization}} \bibinfo{volume}{72}, \bibinfo{number}{2} (\bibinfo{year}{2018}), \bibinfo{pages}{181--217}.
\newblock


\bibitem[Etherscan(2020a)]%
        {bZx1}
\bibfield{author}{\bibinfo{person}{Etherscan}.} \bibinfo{year}{2020}\natexlab{a}.
\newblock \bibinfo{title}{bZx1 Attack Transaction}.
\newblock \bibinfo{howpublished}{\url{https://etherscan.io/tx/0xb5c8bd9430b6cc87a0e2fe110ece6bf527fa4f170a4bc8cd032f768fc5219838}}.
\newblock


\bibitem[Etherscan(2020b)]%
        {Cheesebank}
\bibfield{author}{\bibinfo{person}{Etherscan}.} \bibinfo{year}{2020}\natexlab{b}.
\newblock \bibinfo{title}{Cheesebank Attack Transaction}.
\newblock \bibinfo{howpublished}{\url{https://etherscan.io/tx/0x600a869aa3a259158310a233b815ff67ca41eab8961a49918c2031297a02f1cc}}.
\newblock


\bibitem[Etherscan(2020c)]%
        {Eminence}
\bibfield{author}{\bibinfo{person}{Etherscan}.} \bibinfo{year}{2020}\natexlab{c}.
\newblock \bibinfo{title}{Eminence Attack Transaction}.
\newblock \bibinfo{howpublished}{\url{https://etherscan.io/tx/0x3503253131644dd9f52802d071de74e456570374d586ddd640159cf6fb9b8ad8}}.
\newblock


\bibitem[Etherscan(2020d)]%
        {HarvestUSDC}
\bibfield{author}{\bibinfo{person}{Etherscan}.} \bibinfo{year}{2020}\natexlab{d}.
\newblock \bibinfo{title}{Harvest\_USDC Attack Transaction}.
\newblock \bibinfo{howpublished}{\url{https://etherscan.io/tx/0x35f8d2f572fceaac9288e5d462117850ef2694786992a8c3f6d02612277b0877}}.
\newblock


\bibitem[Etherscan(2020e)]%
        {HarvestUSDT}
\bibfield{author}{\bibinfo{person}{Etherscan}.} \bibinfo{year}{2020}\natexlab{e}.
\newblock \bibinfo{title}{Harvest\_USDT Attack Transaction}.
\newblock \bibinfo{howpublished}{\url{https://etherscan.io/tx/0x0fc6d2ca064fc841bc9b1c1fad1fbb97bcea5c9a1b2b66ef837f1227e06519a6}}.
\newblock


\bibitem[Etherscan(2020f)]%
        {inverseFi}
\bibfield{author}{\bibinfo{person}{Etherscan}.} \bibinfo{year}{2020}\natexlab{f}.
\newblock \bibinfo{title}{inverseFi Attack Transaction}.
\newblock \bibinfo{howpublished}{\url{https://etherscan.io/tx/0x958236266991bc3fe3b77feaacea120f172c0708ad01c7a715b255f218f9313c}}.
\newblock


\bibitem[Etherscan(2020g)]%
        {ValueDeFi}
\bibfield{author}{\bibinfo{person}{Etherscan}.} \bibinfo{year}{2020}\natexlab{g}.
\newblock \bibinfo{title}{ValueDeFi Attack Transaction}.
\newblock \bibinfo{howpublished}{\url{https://etherscan.io/tx/0x46a03488247425f845e444b9c10b52ba3c14927c687d38287c0faddc7471150a}}.
\newblock


\bibitem[Etherscan(2020h)]%
        {Warp}
\bibfield{author}{\bibinfo{person}{Etherscan}.} \bibinfo{year}{2020}\natexlab{h}.
\newblock \bibinfo{title}{Warp Attack Transaction}.
\newblock \bibinfo{howpublished}{\url{https://etherscan.io/tx/0x8bb8dc5c7c830bac85fa48acad2505e9300a91c3ff239c9517d0cae33b595090}}.
\newblock


\bibitem[Etherscan(2021)]%
        {Yearn}
\bibfield{author}{\bibinfo{person}{Etherscan}.} \bibinfo{year}{2021}\natexlab{}.
\newblock \bibinfo{title}{Yearn Attack Transaction}.
\newblock \bibinfo{howpublished}{\url{https://etherscan.io/tx/0xf6022012b73770e7e2177129e648980a82aab555f9ac88b8a9cda3ec44b30779}}.
\newblock


\bibitem[Etherscan(2023)]%
        {etherscan}
\bibfield{author}{\bibinfo{person}{Etherscan}.} \bibinfo{year}{2023}\natexlab{}.
\newblock \bibinfo{title}{The Ethereum Blockchain Explorer}.
\newblock
\newblock
\urldef\tempurl%
\url{https://etherscan.io/}
\showURL{%
\tempurl}


\bibitem[Feist et~al\mbox{.}(2019)]%
        {feist2019slither}
\bibfield{author}{\bibinfo{person}{Josselin Feist}, \bibinfo{person}{Gustavo Grieco}, {and} \bibinfo{person}{Alex Groce}.} \bibinfo{year}{2019}\natexlab{}.
\newblock \showarticletitle{Slither: a static analysis framework for smart contracts}. In \bibinfo{booktitle}{\emph{2019 IEEE/ACM 2nd International Workshop on Emerging Trends in Software Engineering for Blockchain (WETSEB)}}. IEEE, \bibinfo{pages}{8--15}.
\newblock


\bibitem[Feng et~al\mbox{.}(2019)]%
        {feng2019precise}
\bibfield{author}{\bibinfo{person}{Yu Feng}, \bibinfo{person}{Emina Torlak}, {and} \bibinfo{person}{Rastislav Bodik}.} \bibinfo{year}{2019}\natexlab{}.
\newblock \showarticletitle{Precise attack synthesis for smart contracts}.
\newblock \bibinfo{journal}{\emph{arXiv preprint arXiv:1902.06067}} (\bibinfo{year}{2019}).
\newblock


\bibitem[Finance(2022)]%
        {curvefi}
\bibfield{author}{\bibinfo{person}{Curve Finance}.} \bibinfo{year}{2022}\natexlab{}.
\newblock \bibinfo{title}{Curve Finance}.
\newblock \bibinfo{howpublished}{\url{https://curve.fi/}}.
\newblock


\bibitem[Foundation(2023a)]%
        {solidity}
\bibfield{author}{\bibinfo{person}{Ethereum Foundation}.} \bibinfo{year}{2023}\natexlab{a}.
\newblock \bibinfo{title}{Solidity Programming Language}.
\newblock \bibinfo{howpublished}{\url{https://docs.soliditylang.org/en/v0.8.14/}}.
\newblock
\newblock
\shownote{Accessed: 2023-03-24}.


\bibitem[Foundation(2023b)]%
        {vyper}
\bibfield{author}{\bibinfo{person}{Ethereum Foundation}.} \bibinfo{year}{2023}\natexlab{b}.
\newblock \bibinfo{title}{Vyper Programming Language}.
\newblock \bibinfo{howpublished}{\url{https://vyper.readthedocs.io/en/stable/}}.
\newblock
\newblock
\shownote{Accessed: 2023-02-24}.


\bibitem[{Foundry Contributors}(2023)]%
        {Foundry}
\bibfield{author}{\bibinfo{person}{{Foundry Contributors}}.} \bibinfo{year}{2023}\natexlab{}.
\newblock \bibinfo{title}{Foundry}.
\newblock \bibinfo{howpublished}{\url{https://github.com/foundry-rs/foundry/}}.
\newblock
\newblock
\shownote{Accessed: 2023-01-31}.


\bibitem[Frank et~al\mbox{.}(2020)]%
        {frank2020ethbmc}
\bibfield{author}{\bibinfo{person}{Joel Frank}, \bibinfo{person}{Cornelius Aschermann}, {and} \bibinfo{person}{Thorsten Holz}.} \bibinfo{year}{2020}\natexlab{}.
\newblock \showarticletitle{$\{$ETHBMC$\}$: A Bounded Model Checker for Smart Contracts}. In \bibinfo{booktitle}{\emph{29th USENIX Security Symposium (USENIX Security 20)}}. \bibinfo{pages}{2757--2774}.
\newblock


\bibitem[FtmScan(2022)]%
        {OneRing}
\bibfield{author}{\bibinfo{person}{FtmScan}.} \bibinfo{year}{2022}\natexlab{}.
\newblock \bibinfo{title}{OneRing Attack Transaction}.
\newblock \bibinfo{howpublished}{\url{https://ftmscan.com/tx/0xca8dd33850e29cf138c8382e17a19e77d7331b57c7a8451648788bbb26a70145}}.
\newblock


\bibitem[FTMScan(2023)]%
        {FTMScan}
\bibfield{author}{\bibinfo{person}{FTMScan}.} \bibinfo{year}{2023}\natexlab{}.
\newblock \bibinfo{title}{The Fantom Blockchain Explorer}.
\newblock
\newblock
\urldef\tempurl%
\url{https://ftmscan.com/}
\showURL{%
\tempurl}


\bibitem[Gan et~al\mbox{.}(2022)]%
        {gan2022understanding}
\bibfield{author}{\bibinfo{person}{Rundong Gan}, \bibinfo{person}{Le Wang}, \bibinfo{person}{Xiangyu Ruan}, {and} \bibinfo{person}{Xiaodong Lin}.} \bibinfo{year}{2022}\natexlab{}.
\newblock \showarticletitle{Understanding Flash-Loan-based Wash Trading}. In \bibinfo{booktitle}{\emph{Proceedings of the 4th ACM Conference on Advances in Financial Technologies}}. \bibinfo{pages}{74--88}.
\newblock


\bibitem[Ghaleb et~al\mbox{.}(2023)]%
        {ghaleb2023achecker}
\bibfield{author}{\bibinfo{person}{Asem Ghaleb}, \bibinfo{person}{Julia Rubin}, {and} \bibinfo{person}{Karthik Pattabiraman}.} \bibinfo{year}{2023}\natexlab{}.
\newblock \showarticletitle{AChecker: Statically Detecting Smart Contract Access Control Vulnerabilities}.
\newblock \bibinfo{journal}{\emph{Proc. ACM ICSE}} (\bibinfo{year}{2023}).
\newblock


\bibitem[Grieco et~al\mbox{.}(2020)]%
        {grieco2020echidna}
\bibfield{author}{\bibinfo{person}{Gustavo Grieco}, \bibinfo{person}{Will Song}, \bibinfo{person}{Artur Cygan}, \bibinfo{person}{Josselin Feist}, {and} \bibinfo{person}{Alex Groce}.} \bibinfo{year}{2020}\natexlab{}.
\newblock \showarticletitle{Echidna: effective, usable, and fast fuzzing for smart contracts}. In \bibinfo{booktitle}{\emph{Proceedings of the 29th ACM SIGSOFT International Symposium on Software Testing and Analysis}}. \bibinfo{pages}{557--560}.
\newblock


\bibitem[Jiang et~al\mbox{.}(2018)]%
        {jiang2018contractfuzzer}
\bibfield{author}{\bibinfo{person}{Bo Jiang}, \bibinfo{person}{Ye Liu}, {and} \bibinfo{person}{Wing~Kwong Chan}.} \bibinfo{year}{2018}\natexlab{}.
\newblock \showarticletitle{Contractfuzzer: Fuzzing smart contracts for vulnerability detection}. In \bibinfo{booktitle}{\emph{2018 33rd IEEE/ACM International Conference on Automated Software Engineering (ASE)}}. IEEE, \bibinfo{pages}{259--269}.
\newblock


\bibitem[Joe and Kuo(2008)]%
        {joe2008constructing}
\bibfield{author}{\bibinfo{person}{Stephen Joe} {and} \bibinfo{person}{Frances~Y Kuo}.} \bibinfo{year}{2008}\natexlab{}.
\newblock \showarticletitle{Constructing Sobol sequences with better two-dimensional projections}.
\newblock \bibinfo{journal}{\emph{SIAM Journal on Scientific Computing}} \bibinfo{volume}{30}, \bibinfo{number}{5} (\bibinfo{year}{2008}), \bibinfo{pages}{2635--2654}.
\newblock


\bibitem[Kalra et~al\mbox{.}(2018)]%
        {kalra2018zeus}
\bibfield{author}{\bibinfo{person}{Sukrit Kalra}, \bibinfo{person}{Seep Goel}, \bibinfo{person}{Mohan Dhawan}, {and} \bibinfo{person}{Subodh Sharma}.} \bibinfo{year}{2018}\natexlab{}.
\newblock \showarticletitle{Zeus: analyzing safety of smart contracts.}. In \bibinfo{booktitle}{\emph{Ndss}}. \bibinfo{pages}{1--12}.
\newblock


\bibitem[King(1976)]%
        {king1976symbolic}
\bibfield{author}{\bibinfo{person}{James~C King}.} \bibinfo{year}{1976}\natexlab{}.
\newblock \showarticletitle{Symbolic execution and program testing}.
\newblock \bibinfo{journal}{\emph{Commun. ACM}} \bibinfo{volume}{19}, \bibinfo{number}{7} (\bibinfo{year}{1976}), \bibinfo{pages}{385--394}.
\newblock


\bibitem[Liu et~al\mbox{.}(2022)]%
        {liu2022finding}
\bibfield{author}{\bibinfo{person}{Ye Liu}, \bibinfo{person}{Yi Li}, \bibinfo{person}{Shang-Wei Lin}, {and} \bibinfo{person}{Cyrille Artho}.} \bibinfo{year}{2022}\natexlab{}.
\newblock \showarticletitle{Finding permission bugs in smart contracts with role mining}. In \bibinfo{booktitle}{\emph{Proceedings of the 31st ACM SIGSOFT International Symposium on Software Testing and Analysis}}. \bibinfo{pages}{716--727}.
\newblock


\bibitem[Liu et~al\mbox{.}(2020)]%
        {liu2020towards}
\bibfield{author}{\bibinfo{person}{Ye Liu}, \bibinfo{person}{Yi Li}, \bibinfo{person}{Shang-Wei Lin}, {and} \bibinfo{person}{Rong Zhao}.} \bibinfo{year}{2020}\natexlab{}.
\newblock \showarticletitle{Towards automated verification of smart contract fairness}. In \bibinfo{booktitle}{\emph{Proceedings of the 28th ACM Joint Meeting on European Software Engineering Conference and Symposium on the Foundations of Software Engineering}}. \bibinfo{pages}{666--677}.
\newblock


\bibitem[Luu et~al\mbox{.}(2016)]%
        {luu2016making}
\bibfield{author}{\bibinfo{person}{Loi Luu}, \bibinfo{person}{Duc-Hiep Chu}, \bibinfo{person}{Hrishi Olickel}, \bibinfo{person}{Prateek Saxena}, {and} \bibinfo{person}{Aquinas Hobor}.} \bibinfo{year}{2016}\natexlab{}.
\newblock \showarticletitle{Making smart contracts smarter}. In \bibinfo{booktitle}{\emph{Proceedings of the 2016 ACM SIGSAC conference on computer and communications security}}. \bibinfo{pages}{254--269}.
\newblock


\bibitem[Martin~Derka(2023)]%
        {adoption}
\bibfield{author}{\bibinfo{person}{Quantstamp Martin~Derka}.} \bibinfo{year}{2023}\natexlab{}.
\newblock \bibinfo{title}{Automated Flash Loan Attack Synthesis}.
\newblock \bibinfo{howpublished}{\url{https://www.youtube.com/watch?v=e6VR7Rv-jiY&t=100s&ab_channel=DeFiSecuritySummit}}.
\newblock


\bibitem[McKay(2022)]%
        {mckay2022defi}
\bibfield{author}{\bibinfo{person}{Jack McKay}.} \bibinfo{year}{2022}\natexlab{}.
\newblock \showarticletitle{DeFi-ing Cyber Attacks}.
\newblock  (\bibinfo{year}{2022}).
\newblock


\bibitem[Montgomery et~al\mbox{.}(2021)]%
        {montgomery2021introduction}
\bibfield{author}{\bibinfo{person}{Douglas~C Montgomery}, \bibinfo{person}{Elizabeth~A Peck}, {and} \bibinfo{person}{G~Geoffrey Vining}.} \bibinfo{year}{2021}\natexlab{}.
\newblock \bibinfo{booktitle}{\emph{Introduction to linear regression analysis}}.
\newblock \bibinfo{publisher}{John Wiley \& Sons}.
\newblock


\bibitem[Mossberg et~al\mbox{.}(2019)]%
        {mossberg2019manticore}
\bibfield{author}{\bibinfo{person}{Mark Mossberg}, \bibinfo{person}{Felipe Manzano}, \bibinfo{person}{Eric Hennenfent}, \bibinfo{person}{Alex Groce}, \bibinfo{person}{Gustavo Grieco}, \bibinfo{person}{Josselin Feist}, \bibinfo{person}{Trent Brunson}, {and} \bibinfo{person}{Artem Dinaburg}.} \bibinfo{year}{2019}\natexlab{}.
\newblock \showarticletitle{Manticore: A user-friendly symbolic execution framework for binaries and smart contracts}. In \bibinfo{booktitle}{\emph{2019 34th IEEE/ACM International Conference on Automated Software Engineering (ASE)}}. IEEE, \bibinfo{pages}{1186--1189}.
\newblock


\bibitem[Nguyen et~al\mbox{.}(2020)]%
        {nguyen2020sfuzz}
\bibfield{author}{\bibinfo{person}{Tai~D Nguyen}, \bibinfo{person}{Long~H Pham}, \bibinfo{person}{Jun Sun}, \bibinfo{person}{Yun Lin}, {and} \bibinfo{person}{Quang~Tran Minh}.} \bibinfo{year}{2020}\natexlab{}.
\newblock \showarticletitle{sfuzz: An efficient adaptive fuzzer for solidity smart contracts}. In \bibinfo{booktitle}{\emph{Proceedings of the ACM/IEEE 42nd International Conference on Software Engineering}}. \bibinfo{pages}{778--788}.
\newblock


\bibitem[Pedregosa et~al\mbox{.}(2011)]%
        {scikit-learn}
\bibfield{author}{\bibinfo{person}{F. Pedregosa}, \bibinfo{person}{G. Varoquaux}, \bibinfo{person}{A. Gramfort}, \bibinfo{person}{V. Michel}, \bibinfo{person}{B. Thirion}, \bibinfo{person}{O. Grisel}, \bibinfo{person}{M. Blondel}, \bibinfo{person}{P. Prettenhofer}, \bibinfo{person}{R. Weiss}, \bibinfo{person}{V. Dubourg}, \bibinfo{person}{J. Vanderplas}, \bibinfo{person}{A. Passos}, \bibinfo{person}{D. Cournapeau}, \bibinfo{person}{M. Brucher}, \bibinfo{person}{M. Perrot}, {and} \bibinfo{person}{E. Duchesnay}.} \bibinfo{year}{2011}\natexlab{}.
\newblock \showarticletitle{Scikit-learn: Machine Learning in {P}ython}.
\newblock \bibinfo{journal}{\emph{Journal of Machine Learning Research}}  \bibinfo{volume}{12} (\bibinfo{year}{2011}), \bibinfo{pages}{2825--2830}.
\newblock


\bibitem[Qin et~al\mbox{.}(2021)]%
        {qin2021attacking}
\bibfield{author}{\bibinfo{person}{Kaihua Qin}, \bibinfo{person}{Liyi Zhou}, \bibinfo{person}{Benjamin Livshits}, {and} \bibinfo{person}{Arthur Gervais}.} \bibinfo{year}{2021}\natexlab{}.
\newblock \showarticletitle{Attacking the defi ecosystem with flash loans for fun and profit}. In \bibinfo{booktitle}{\emph{International Conference on Financial Cryptography and Data Security}}. Springer, \bibinfo{pages}{3--32}.
\newblock


\bibitem[Quantstamp(2023a)]%
        {adoptionNews1}
\bibfield{author}{\bibinfo{person}{Quantstamp}.} \bibinfo{year}{2023}\natexlab{a}.
\newblock \bibinfo{title}{Economic Exploit Analysis}.
\newblock \bibinfo{howpublished}{\url{https://quantstamp.com/economic-exploits}}.
\newblock


\bibitem[Quantstamp(2023b)]%
        {adoptionNews2}
\bibfield{author}{\bibinfo{person}{Quantstamp}.} \bibinfo{year}{2023}\natexlab{b}.
\newblock \bibinfo{title}{Web3 Security Firm Quantstamp Launches Novel Economic Exploit Analysis Service to Combat Flash Loan Attacks}.
\newblock \bibinfo{howpublished}{\url{https://www.newswire.ca/news-releases/web3-security-firm-quantstamp-launches-novel-economic-exploit-analysis-service-to-combat-flash-loan-attacks-833815039.html}}.
\newblock


\bibitem[Ramezany et~al\mbox{.}(2023)]%
        {ramezany2023midnight}
\bibfield{author}{\bibinfo{person}{Shahin Ramezany}, \bibinfo{person}{Rachsuda Setthawong}, {and} \bibinfo{person}{Pisal Setthawong}.} \bibinfo{year}{2023}\natexlab{}.
\newblock \showarticletitle{Midnight: An Efficient Event-driven EVM Transaction Security Monitoring Approach For Flash Loan Detection}. In \bibinfo{booktitle}{\emph{2023 20th International Joint Conference on Computer Science and Software Engineering (JCSSE)}}. IEEE, \bibinfo{pages}{237--241}.
\newblock


\bibitem[Rekt(2020)]%
        {Rekt}
\bibfield{author}{\bibinfo{person}{Rekt}.} \bibinfo{year}{2020}\natexlab{}.
\newblock \bibinfo{title}{WARP FINANCE - REKT}.
\newblock \bibinfo{howpublished}{\url{https://rekt.news/warp-finance-rekt/}}.
\newblock


\bibitem[Rukundo and Cao(2012)]%
        {rukundo2012nearest}
\bibfield{author}{\bibinfo{person}{Olivier Rukundo} {and} \bibinfo{person}{Hanqiang Cao}.} \bibinfo{year}{2012}\natexlab{}.
\newblock \showarticletitle{Nearest neighbor value interpolation}.
\newblock \bibinfo{journal}{\emph{arXiv preprint arXiv:1211.1768}} (\bibinfo{year}{2012}).
\newblock


\bibitem[Rush(2023)]%
        {trueblocks}
\bibfield{author}{\bibinfo{person}{Thomas~Jay Rush}.} \bibinfo{year}{2023}\natexlab{}.
\newblock \bibinfo{title}{TrueBlocks}.
\newblock \bibinfo{howpublished}{\url{https://trueblocks.io/}}.
\newblock


\bibitem[Shou et~al\mbox{.}(2023)]%
        {shou2023ityfuzz}
\bibfield{author}{\bibinfo{person}{Chaofan Shou}, \bibinfo{person}{Shangyin Tan}, {and} \bibinfo{person}{Koushik Sen}.} \bibinfo{year}{2023}\natexlab{}.
\newblock \showarticletitle{ItyFuzz: Snapshot-Based Fuzzer for Smart Contract}. In \bibinfo{booktitle}{\emph{Proceedings of the 32nd ACM SIGSOFT International Symposium on Software Testing and Analysis}}. \bibinfo{pages}{322--333}.
\newblock


\bibitem[SlowMist(2021)]%
        {WarpFiAnalysis}
\bibfield{author}{\bibinfo{person}{SlowMist}.} \bibinfo{year}{2021}\natexlab{}.
\newblock \bibinfo{title}{Analysis of Warp Finance Hacked Incident}.
\newblock \bibinfo{howpublished}{\url{https://slowmist.medium.com/analysis-of-warp-finance-hacked-incident-cb12a1af74cc}}.
\newblock


\bibitem[SlowMist(2023)]%
        {SlowMistStats}
\bibfield{author}{\bibinfo{person}{SlowMist}.} \bibinfo{year}{2023}\natexlab{}.
\newblock \bibinfo{title}{The 10 most common attacks}.
\newblock \bibinfo{howpublished}{\url{https://hacked.slowmist.io/en/statistics/?c=all&d=all}}.
\newblock
\newblock
\shownote{Accessed: 2023-02-01}.


\bibitem[Swap(2022)]%
        {pancake}
\bibfield{author}{\bibinfo{person}{Pancake Swap}.} \bibinfo{year}{2022}\natexlab{}.
\newblock \bibinfo{title}{Pancake Swap}.
\newblock \bibinfo{howpublished}{\url{https://pancakeswap.finance/}}.
\newblock


\bibitem[Tikhomirov et~al\mbox{.}(2018)]%
        {tikhomirov2018smartcheck}
\bibfield{author}{\bibinfo{person}{Sergei Tikhomirov}, \bibinfo{person}{Ekaterina Voskresenskaya}, \bibinfo{person}{Ivan Ivanitskiy}, \bibinfo{person}{Ramil Takhaviev}, \bibinfo{person}{Evgeny Marchenko}, {and} \bibinfo{person}{Yaroslav Alexandrov}.} \bibinfo{year}{2018}\natexlab{}.
\newblock \showarticletitle{Smartcheck: Static analysis of ethereum smart contracts}. In \bibinfo{booktitle}{\emph{Proceedings of the 1st International Workshop on Emerging Trends in Software Engineering for Blockchain}}. \bibinfo{pages}{9--16}.
\newblock


\bibitem[Tsankov et~al\mbox{.}(2018)]%
        {tsankov2018securify}
\bibfield{author}{\bibinfo{person}{Petar Tsankov}, \bibinfo{person}{Andrei Dan}, \bibinfo{person}{Dana Drachsler-Cohen}, \bibinfo{person}{Arthur Gervais}, \bibinfo{person}{Florian Buenzli}, {and} \bibinfo{person}{Martin Vechev}.} \bibinfo{year}{2018}\natexlab{}.
\newblock \showarticletitle{Securify: Practical security analysis of smart contracts}. In \bibinfo{booktitle}{\emph{Proceedings of the 2018 ACM SIGSAC Conference on Computer and Communications Security}}. \bibinfo{pages}{67--82}.
\newblock


\bibitem[{Uniswap Labs}(2023)]%
        {Uniswap}
\bibfield{author}{\bibinfo{person}{{Uniswap Labs}}.} \bibinfo{year}{2023}\natexlab{}.
\newblock \bibinfo{title}{{Uniswap Protocol}}.
\newblock \bibinfo{howpublished}{\url{https://uniswap.org/}}.
\newblock
\newblock
\shownote{Accessed: 2023-11-08}.


\bibitem[Virtanen et~al\mbox{.}(2020)]%
        {2020SciPy-NMeth}
\bibfield{author}{\bibinfo{person}{Pauli Virtanen}, \bibinfo{person}{Ralf Gommers}, \bibinfo{person}{Travis~E. Oliphant}, \bibinfo{person}{Matt Haberland}, \bibinfo{person}{Tyler Reddy}, \bibinfo{person}{David Cournapeau}, \bibinfo{person}{Evgeni Burovski}, \bibinfo{person}{Pearu Peterson}, \bibinfo{person}{Warren Weckesser}, \bibinfo{person}{Jonathan Bright}, \bibinfo{person}{St{\'e}fan~J. {van der Walt}}, \bibinfo{person}{Matthew Brett}, \bibinfo{person}{Joshua Wilson}, \bibinfo{person}{K.~Jarrod Millman}, \bibinfo{person}{Nikolay Mayorov}, \bibinfo{person}{Andrew R.~J. Nelson}, \bibinfo{person}{Eric Jones}, \bibinfo{person}{Robert Kern}, \bibinfo{person}{Eric Larson}, \bibinfo{person}{C~J Carey}, \bibinfo{person}{{\.I}lhan Polat}, \bibinfo{person}{Yu Feng}, \bibinfo{person}{Eric~W. Moore}, \bibinfo{person}{Jake {VanderPlas}}, \bibinfo{person}{Denis Laxalde}, \bibinfo{person}{Josef Perktold}, \bibinfo{person}{Robert Cimrman}, \bibinfo{person}{Ian Henriksen}, \bibinfo{person}{E.~A. Quintero}, \bibinfo{person}{Charles~R. Harris}, \bibinfo{person}{Anne~M. Archibald}, \bibinfo{person}{Ant{\^o}nio~H. Ribeiro}, \bibinfo{person}{Fabian Pedregosa}, \bibinfo{person}{Paul {van Mulbregt}}, {and} \bibinfo{person}{{SciPy 1.0 Contributors}}.} \bibinfo{year}{2020}\natexlab{}.
\newblock \showarticletitle{{{SciPy} 1.0: Fundamental Algorithms for Scientific Computing in Python}}.
\newblock \bibinfo{journal}{\emph{Nature Methods}}  \bibinfo{volume}{17} (\bibinfo{year}{2020}), \bibinfo{pages}{261--272}.
\newblock
\urldef\tempurl%
\url{https://doi.org/10.1038/s41592-019-0686-2}
\showDOI{\tempurl}


\bibitem[Wales(2015)]%
        {wales2015perspective}
\bibfield{author}{\bibinfo{person}{DJ Wales}.} \bibinfo{year}{2015}\natexlab{}.
\newblock \showarticletitle{Perspective: Insight into reaction coordinates and dynamics from the potential energy landscape}.
\newblock \bibinfo{journal}{\emph{The Journal of chemical physics}} \bibinfo{volume}{142}, \bibinfo{number}{13} (\bibinfo{year}{2015}), \bibinfo{pages}{130901}.
\newblock


\bibitem[Wang et~al\mbox{.}(2022)]%
        {wang2022defiscanner}
\bibfield{author}{\bibinfo{person}{Bin Wang}, \bibinfo{person}{Xiaohan Yuan}, \bibinfo{person}{Li Duan}, \bibinfo{person}{Hongliang Ma}, \bibinfo{person}{Chunhua Su}, {and} \bibinfo{person}{Wei Wang}.} \bibinfo{year}{2022}\natexlab{}.
\newblock \showarticletitle{DeFiScanner: Spotting DeFi Attacks Exploiting Logic Vulnerabilities on Blockchain}.
\newblock \bibinfo{journal}{\emph{IEEE Transactions on Computational Social Systems}} (\bibinfo{year}{2022}).
\newblock


\bibitem[Wang et~al\mbox{.}(2021)]%
        {wang2021towards}
\bibfield{author}{\bibinfo{person}{Dabao Wang}, \bibinfo{person}{Siwei Wu}, \bibinfo{person}{Ziling Lin}, \bibinfo{person}{Lei Wu}, \bibinfo{person}{Xingliang Yuan}, \bibinfo{person}{Yajin Zhou}, \bibinfo{person}{Haoyu Wang}, {and} \bibinfo{person}{Kui Ren}.} \bibinfo{year}{2021}\natexlab{}.
\newblock \showarticletitle{Towards a first step to understand flash loan and its applications in defi ecosystem}. In \bibinfo{booktitle}{\emph{Proceedings of the Ninth International Workshop on Security in Blockchain and Cloud Computing}}. \bibinfo{pages}{23--28}.
\newblock


\bibitem[Wang et~al\mbox{.}(2020)]%
        {wang2020oracle}
\bibfield{author}{\bibinfo{person}{Haijun Wang}, \bibinfo{person}{Ye Liu}, \bibinfo{person}{Yi Li}, \bibinfo{person}{Shang-Wei Lin}, \bibinfo{person}{Cyrille Artho}, \bibinfo{person}{Lei Ma}, {and} \bibinfo{person}{Yang Liu}.} \bibinfo{year}{2020}\natexlab{}.
\newblock \showarticletitle{Oracle-supported dynamic exploit generation for smart contracts}.
\newblock \bibinfo{journal}{\emph{IEEE Transactions on Dependable and Secure Computing}} (\bibinfo{year}{2020}).
\newblock


\bibitem[Werapun et~al\mbox{.}(2022)]%
        {werapun2022flash}
\bibfield{author}{\bibinfo{person}{Warodom Werapun}, \bibinfo{person}{Tanakorn Karode}, \bibinfo{person}{Tanwa Arpornthip}, \bibinfo{person}{Jakapan Suaboot}, \bibinfo{person}{Esther Sangiamkul}, {and} \bibinfo{person}{Pawita Boonrat}.} \bibinfo{year}{2022}\natexlab{}.
\newblock \showarticletitle{The Flash Loan Attack Analysis (FAA) Framework—A Case Study of the Warp Finance Exploitation}. In \bibinfo{booktitle}{\emph{Informatics}}, Vol.~\bibinfo{volume}{10}. MDPI, \bibinfo{pages}{3}.
\newblock


\bibitem[Wood(2012)]%
        {ethereum22}
\bibfield{author}{\bibinfo{person}{Gavin Wood}.} \bibinfo{year}{2012}\natexlab{}.
\newblock \bibinfo{title}{Ethereum: A Secure Decentralised Generalised Transaction Ledger}.
\newblock
\newblock
\newblock
\shownote{\url{https://ethereum.github.io/yellowpaper/paper.pdf}}.


\bibitem[W{\"u}st and Gervais(2018)]%
        {wust2018you}
\bibfield{author}{\bibinfo{person}{Karl W{\"u}st} {and} \bibinfo{person}{Arthur Gervais}.} \bibinfo{year}{2018}\natexlab{}.
\newblock \showarticletitle{Do you need a blockchain?}. In \bibinfo{booktitle}{\emph{2018 Crypto Valley Conference on Blockchain Technology (CVCBT)}}. IEEE, \bibinfo{pages}{45--54}.
\newblock


\bibitem[Xia et~al\mbox{.}(2021)]%
        {xia2021trade}
\bibfield{author}{\bibinfo{person}{Pengcheng Xia}, \bibinfo{person}{Haoyu Wang}, \bibinfo{person}{Bingyu Gao}, \bibinfo{person}{Weihang Su}, \bibinfo{person}{Zhou Yu}, \bibinfo{person}{Xiapu Luo}, \bibinfo{person}{Chao Zhang}, \bibinfo{person}{Xusheng Xiao}, {and} \bibinfo{person}{Guoai Xu}.} \bibinfo{year}{2021}\natexlab{}.
\newblock \showarticletitle{Trade or Trick? Detecting and Characterizing Scam Tokens on Uniswap Decentralized Exchange}.
\newblock \bibinfo{journal}{\emph{Proceedings of the ACM on Measurement and Analysis of Computing Systems}} \bibinfo{volume}{5}, \bibinfo{number}{3} (\bibinfo{year}{2021}), \bibinfo{pages}{1--26}.
\newblock


\bibitem[Xia et~al\mbox{.}(2023)]%
        {xia2023detecting}
\bibfield{author}{\bibinfo{person}{Qing Xia}, \bibinfo{person}{Zhirong Huang}, \bibinfo{person}{Wensheng Dou}, \bibinfo{person}{Yafeng Zhang}, \bibinfo{person}{Fengjun Zhang}, \bibinfo{person}{Geng Liang}, {and} \bibinfo{person}{Chun Zuo}.} \bibinfo{year}{2023}\natexlab{}.
\newblock \showarticletitle{Detecting Flash Loan Based Attacks in Ethereum}. In \bibinfo{booktitle}{\emph{2023 IEEE 43rd International Conference on Distributed Computing Systems (ICDCS)}}. IEEE, \bibinfo{pages}{154--165}.
\newblock


\bibitem[Xu et~al\mbox{.}(2021)]%
        {xu2021sok}
\bibfield{author}{\bibinfo{person}{Jiahua Xu}, \bibinfo{person}{Krzysztof Paruch}, \bibinfo{person}{Simon Cousaert}, {and} \bibinfo{person}{Yebo Feng}.} \bibinfo{year}{2021}\natexlab{}.
\newblock \showarticletitle{Sok: Decentralized exchanges (dex) with automated market maker (AMM) protocols}.
\newblock \bibinfo{journal}{\emph{arXiv preprint arXiv:2103.12732}} (\bibinfo{year}{2021}).
\newblock


\bibitem[Zhang et~al\mbox{.}(2023)]%
        {zhang2023demystifying}
\bibfield{author}{\bibinfo{person}{Zhuo Zhang}, \bibinfo{person}{Brian Zhang}, \bibinfo{person}{Wen Xu}, {and} \bibinfo{person}{Zhiqiang Lin}.} \bibinfo{year}{2023}\natexlab{}.
\newblock \showarticletitle{Demystifying Exploitable Bugs in Smart Contracts}. ICSE.
\newblock


\bibitem[Zheng et~al\mbox{.}(2022)]%
        {zheng2022park}
\bibfield{author}{\bibinfo{person}{Peilin Zheng}, \bibinfo{person}{Zibin Zheng}, {and} \bibinfo{person}{Xiapu Luo}.} \bibinfo{year}{2022}\natexlab{}.
\newblock \showarticletitle{Park: accelerating smart contract vulnerability detection via parallel-fork symbolic execution}. In \bibinfo{booktitle}{\emph{Proceedings of the 31st ACM SIGSOFT International Symposium on Software Testing and Analysis}}. \bibinfo{pages}{740--751}.
\newblock


\end{thebibliography}

\appendix
\newpage
~\newpage

\section{Functions of Curve.Fi Y Pool Contract}

To extend the discussion about the complexity of \emph(Curve.Fi Y Pool) contract
in Section 3
, we include another three functions \\
\codeword{exchange\_underlying},
\codeword{\_exchange},
and \codeword{get\_y}
in Figure~\ref{fig:exchange_underlying},
Figure~\ref{fig:exchange},
and Figure~\ref{fig:get_y} respectively.

\codeword{exchange\_underlying} is the public function usually called by users
to exchange \codeword{dx} amount of underlying asset \codeword{i} to \codeword{j}, and
\codeword{min\_dy} represents the minimum quantity expected to receive from the exchange. 
Once \codeword{exchange\_underlying} is called, it will internally execute a private function 
\codeword{\_exchange}, which will call another private \codeword{get\_y}. \codeword{get\_y}
first calls \codeword{get\_D} to calculate a constant $D$ iteratively. Then it calculates 
an updated value of $y$ iteratively based on $D$ and other constants. This iterative Calculation
impedes symbolic execution since it results in too complicated non-linear constraints.

\begin{figure}[H]
    \centering
    \begin{minipage}[h]{0.9\linewidth}
\begin{lstlisting}
function exchange_underlying( int128 i, int128 j,
                              uint dx, uint min_dy) 
{
  uint[] memory rates = _stored_rates();
  uint N_COINS = rates.length;
  uint[] memory precisions = PRECISION_MUL;
  uint rate_i = rates[i] / precisions[i];
  uint rate_j = rates[j] / precisions[j];
  uint dx_ = (dx * PRECISION) / rate_i;
  // key step: call get_y and then call get_D
  uint dy_ = _exchange(i, j, dx_, rates); 
  uint dy = (dy_ * rate_j) / PRECISION;
  assert(dy >= min_dy, "...");
  bool[] memory tethered = TETHERED;
  uint ok = 0;
  if (tethered[i])
    USDT(u_coins[i]).transferFrom(
        msg.sender, address(this), dx);
  else
    assert_modifiable(
      ERC20(u_coins[i]).transferFrom(
        msg.sender, address(this), dx)
    );
  ERC20(u_coins[i]).approve(coins[i], dx);
  yERC20(coins[i]).deposit(dx);
  yERC20(coins[j]).withdraw(dy_);
  dy=ERC20(u_coins[j]).balanceOf(address(this));
  assert(dy >= min_dy, "...");
  if (tethered[j]) 
    USDT(u_coins[j]).transfer(msg.sender, dy);
  else assert_modifiable(ERC20(u_coins[j])
    .transfer(msg.sender, dy));
  log.TokenExchangeUnderlying(msg.sender, i, dx, j, dy);
}
\end{lstlisting}
    \end{minipage}
    \caption{Source code of exchange\_underlying(). Original code is in Vyper, rewritten in Solidity.}
    \label{fig:exchange_underlying}
    \vspace{-2mm}
\end{figure}

\begin{figure}[H]
    \centering
    \begin{minipage}[h]{0.9\linewidth}
\begin{lstlisting}
function _exchange(int128 i, int128 j,
                   uint dx, uint[] calldata rates) 
{
    uint[N_COINS] xp = _xp(rates);
    uint x = xp[i] + dx * rates[i] / PRECISION;
    uint y = get_y(i, j, x, xp);
    uint dy = xp[j] - y;
    uint dy_fee = dy * self.fee / FEE_DENOMINATOR;
    uint dy_admin_fee = dy_fee * self.admin_fee / FEE_DENOMINATOR;
    balances[i] = x * PRECISION / rates[i];
    balances[j] = (y + (dy_fee - dy_admin_fee)) * PRECISION / rates[j]
    uint _dy = (dy - dy_fee) * PRECISION / rates[j]
    return _dy;
}

\end{lstlisting}
    \end{minipage}
    \caption{Source code of exchange(). Original code is in Vyper, rewritten in Solidity.}
    \label{fig:exchange}
    \vspace{-2mm}
\end{figure}

\begin{figure}[H]
    \centering
    \begin{minipage}[h]{0.9\linewidth}
\begin{lstlisting}
function get_y( int128 i, int128 j,
                uint x, uint[] calldata _xp) returns (uint) 
{ 
  uint N_COINS = _xp.length;
  assert(
    i != j && i >= 0 && j >= 0 && 
    uint(i) < N_COINS && uint(j) < N_COINS
  );
  uint D = get_D(_xp);
  uint c = D;
  uint S_ = 0;
  uint Ann = A * N_COINS; // A is a constant
  uint _x = 0;
  for (uint _i = 0; _i < N_COINS; _i = _i + 1) {
    if (_i == uint(i)) {
      _x = x;
    } else if (_i != uint(j)) {
      _x = _xp[_i];
    } else continue;
    S_ += _x;
    c = (c * D) / (_x * N_COINS);
  }
  c = (c * D) / (Ann * N_COINS);
  uint b = S_ + D / Ann; // - D
  uint y_prev = 0;
  uint y = D;
  for (uint _i = 0; _i < 255; _i = _i + 1) {
    y_prev = y;
    y = (y * y + c) / (2 * y + b - D);
    // Equality with the precision of 1
    if (y > y_prev) {
      if (y - y_prev <= 1) break;
    } else {
      if (y_prev - y <= 1) break;
    }
  }
  return y;
}
\end{lstlisting}
    \end{minipage}
    \caption{Source code of get\_y(). Original code is in Vyper, rewritten in Solidity.}
    \label{fig:get_y}
    \vspace{-2mm}
\end{figure}

\section{Warp Finance Case Study}
\label{subsec:Warp}

On Dec. 18, 2020, Warp Finance suffered a flash loan attack which leads to about \$7.8 million loss, 
according to Rekt~\cite{Rekt}. The first attack happened in 
the transaction~\cite{Warp}. 
A detailed postmortem analysis~\cite{WarpFiAnalysis} is written by SlowMist on Medium.

\noindent \textbf{Attack Root Cause:}
The core of Warp attack is a design flaw of calculating 
the price of LP tokens. The price of LP token is calculated as (amount(WETH) in the pool * WETH price + amount(DAI) in the pool * DAI price) / total supply of LP. Even though the developer uses Uniswap official price oracle to (correctly) calculate the prices of WETH and DAI, they failed to realize the amounts of WETH and DAI can also be manipulated by flash loans. The exploiter took advantage of this point, pumped up the LP price, and falsely borrow excessive USDC and DAI from the pool.

\noindent \textbf{Real Attack Vector in History:}
\\
Action 1: \codeword{mint(2900030e18)} 
\\ 
Action 2: \codeword{swap(WETH, DAI, 341217e18)} 
\\ 
Action 3: \codeword{provideCollateral(94349e18)} 
\\ 
Action 4: \codeword{borrowSC(USDC, 3917983e6)} 
\\
Action 5: \codeword{borrowSC(DAI, 3862646e18)} 
\\
Action 6: \codeword{swap(DAI, WETH, 47622329e18)} 
\\ 
Adjusted profit: ~1,693,523 USD (DAI + USDC + WETH)

First the exploiter flash loaned WETH and DAI and \codeword{mint} 2900030 DAI and equivalent WETH to Uniswap's WETH-DAI pair. Then the attacker swapped a huge amount of WETH into DAI via the pair to increase the total value of the WETH-DAI pool, pumping up the unit price of LP token. The exploiter then mortgages the previously obtained LP Token through the \codeword{provideCollateral} function. As the unit price of LP Token becomes higher, the LP tokens mortgaged by the attacker give an abnormally high borrow limit thus the attacker can borrow more make profit.

\noindent \textbf{Best Attack Vector from {\name}:}
\\
Action 1: \codeword{swap(WETH, DAI, 480069e18)} 
\\
Action 2: \codeword{mint(322007e18)} 
\\
Action 3: \codeword{provideCollateral(65937e18)} 
\\ 
Action 4: \codeword{borrowSC(USDC, 3862530e6)} 
\\
Action 5: \codeword{borrowSC(DAI, 3917914e18)} 
\\
Action 6: \codeword{swap(DAI, WETH, 48862671e18)} 
\\ 
Adjusted profit: ~3,368,718 USD (DAI + USDC + WETH)

It is quite surprising that the attack vector from {\name} gets a higher profit compared to the attacker in history. The reason behind this is the attack vector given by {\name} spends more on manipulating Uniswap pair(Action 1 and Action 6) and spends less on the LP Token mortgaged(Action 2 and Action 3). In this attack, to get the same borrow limit(which can be used to borrow USDC and DAI in Action 4 and Action 5), it is more economical to spend on manipulating the Uniswap pair than collateralizing more LP tokens.

\section{Annotate Action Candidates}
\label{sec:annotation}
As discussed in Section 3, {\name} makes use of annotations of prestates and poststates to 
drive the approximations. This section provides more details about the annotation process.

In principle, \textbf{states} refer to read-only functions or contract variables that contribute to the calculation of token transfers. 
A \textbf{prestate} refers to the state before the execution of an action candidate. If two executions share identical prestates and related inputs, their token transfers must also be the same. {\name} approximates a function which maps (prestates, inputs) to token transfers. Executing an action will also alter states, which are denoted as \textbf{poststates}. Poststates of an action will be the prestates of the next action. {\name} also approximates a function which maps (prestates, inputs) to poststates. By utilizing these two types of approximations, {\name} automatically builds a constrainted optimization problem to search for the optimal parameters.

Here we use the Vault contract in Figure~\ref{fig:depositwithdraw}
as an example to illustrate how we annotate \codeword{deposit} and \codeword{withdraw} functions.

    \begin{figure}[h]
        \centering
        \begin{minipage}[t]{1.0\linewidth}
\begin{lstlisting}
function totalSupply() public view returns (uint256) {
    return _totalSupply;
}

function underlyingBalanceInVault() view public returns (uint256) {
  return IERC20(underlying()).balanceOf(address(this));
}

function underlyingBalanceWithInvestment() view public returns (uint256) {
    if (address(strategy()) == address(0)) {
      return underlyingBalanceInVault();
    }
    return underlyingBalanceInVault()
            .add(IStrategy(strategy()).investedUnderlyingBalance());
}

function deposit(uint256 amount) external {
    _deposit(amount, msg.sender, msg.sender);
}

function _deposit(uint amount, address sender, address beneficiary) internal {
    uint toMint = totalSupply() == 0 ? amount: 
        amount.mul(totalSupply()).div(underlyingBalanceWithInvestment());
    _mint(beneficiary, toMint);
    IERC20(underlying()).safeTransferFrom(sender, address(this), amount);
    emit Deposit(beneficiary, amount);
}

function withdraw(uint numberOfShares) external {
    uint totalSupply = totalSupply();
    _burn(msg.sender, numberOfShares);
    uint underlyingAmountToWithdraw = underlyingBalanceWithInvestment()
        .mul(numberOfShares)
        .div(totalSupply);
    if (underlyingAmountToWithdraw > underlyingBalanceInVault()) {
        if (numberOfShares == totalSupply) {
            IStrategy(strategy()).withdrawAllToVault();
        } else {
            uint missing = underlyingAmountToWithdraw.sub(
                underlyingBalanceInVault()
            );
            IStrategy(strategy()).withdrawToVault(missing);
        }
        // recalculate to improve accuracy
        underlyingAmountToWithdraw = Math.min(
            underlyingBalanceWithInvestment().mul(numberOfShares).div(
                totalSupply
            ),
            underlyingBalanceInVault()
        );
    }
    IERC20(underlying()).safeTransfer(msg.sender, underlyingAmountToWithdraw);
    emit Withdraw(msg.sender, underlyingAmountToWithdraw);
}
\end{lstlisting}
        \end{minipage}
        \vspace{-3mm}
        \caption{The Vault \codeword{deposit} and \codeword{withdraw} methods.}
        \label{fig:depositwithdraw}
    \end{figure}

Function \codeword{deposit} moves two tokens. First, it transfers \texttt{amount} USDC from msg.sender, which is exactly the input argument. No further annotation is needed for this token transfer.
\justify
Second, it mints and transfers \texttt{toMint} fUSDC tokens to the msg.sender. We also find \texttt{toMint} is calculated using the input \texttt{amount} and two functions \texttt{totalSupply()} and \texttt{underlyingBalanceWithInvestment()}.

\texttt{totalSupply()} returns contract variable \texttt{totalSupply} so it is not further expandable. \texttt{underlyingBalanceWithInvestment()} can be further expanded to another two read-only functions \texttt{underlyingBalanceInVault()} and \texttt{investedUnderlyingBalance()}. Both functions are not further expandable because they are external read-only function calls.

\begin{flushleft}
    \hyphenpenalty=10000
    \exhyphenpenalty=10000

It is also observed that \texttt{totalSupply()}, \texttt{underlyingBalanceInVault()} are changed after the execution of \codeword{deposit}, while \texttt{investedUnderlyingBalance()} remains unchanged. Thus, we annotate \codeword{deposit} as follows:

\end{flushleft}

\vspace{5pt}
{
\footnotesize
\fbox{%
\parbox{.97\columnwidth}{
prestates= \{\texttt{totalSupply()}, \texttt{underlyingBalanceInVault()}, \\
        ,\ \ \ \ \ \ \ \ \ \ \ \ \ \ \ \ \  \texttt{investedUnderlyingBalance()}\},  \\
poststates= \{\texttt{totalSupply()}, \texttt{underlyingBalanceInVault()}\},  \\
tokenIn = \{USDC\}, 
tokenOut = \{fUSDC\}. 
}}
}

\justify
Function \codeword{withdraw} also moves two tokens. First, it transfers and burns \texttt{numberOfShares} fUSDC tokens from msg.sender. \texttt{numberOfShares} is just the input argument. 

\begin{flushleft}
    \hyphenpenalty=10000
    \exhyphenpenalty=10000
Second, it transfers \texttt{underlyingAmountToWithdraw} USDC tokens to msg.sender. \texttt{underlyingAmountToWithdraw} is calculated using the input \texttt{numberOfShares} and two states \texttt{totalSupply()} and \texttt{underlyingBalanceWithInvestment()}. Similarly, \texttt{underlyingBalanceWithInvestment()} can be further expanded to \texttt{underlyingBalanceInVault()} and \texttt{investedUnderlyingBalance()}.
\end{flushleft}

\begin{flushleft}
    \hyphenpenalty=10000
    \exhyphenpenalty=10000
We also observe that \texttt{totalSupply()}, \texttt{underlyingBalanceInVault()} are changed after the execution of \codeword{withdraw}, while \texttt{investedUnderlyingBalance()} remains unchanged. Thus, we annotate \codeword{withdraw} as follows:
\end{flushleft}

\vspace{5pt}
{
\footnotesize
\fbox{%
\parbox{.97\columnwidth}{
prestates= \{\texttt{totalSupply()}, \texttt{underlyingBalanceInVault()}, \\
        ,\ \ \ \ \ \ \ \ \ \ \ \ \ \ \ \ \  \texttt{investedUnderlyingBalance()}\},  \\
poststates= \{\texttt{totalSupply()}, \texttt{underlyingBalanceInVault()}\},  \\
tokenIn = \{fUSDC\}, 
tokenOut = \{USDC\}. 
}}
}

In some cases, action A may internally call another function B from a different contract, which ultimately results in token transfers to/from users. In such scenarios, we need to annotate function B as the annotations for action A. Additionally, there may be instances where action C alters the states of another action D, despite not belonging to the same protocol. For instance, action D could utilize action C's contract states as an oracle, and action C may modify these states. In such situations, it is essential to include action D's prestates in action C's annotations. We would then collect data points for these state changes and approximate them separately in the approximation stage.

We believe that the annotating process should not pose significant difficulties for security analysts. Furthermore, we are optimistic that this process can be further automated through either static analysis of (open-source) contracts or dynamic analysis of the EVM level execution traces of deployed contracts (closed-source), and is left as future work.

\section{{\gen}: Action Candidates Identification}
Flash loan attacks typically focus on victim contracts containing functions capable of transferring tokens,\footnote{Here, tokens refer to various forms of DeFi tokens, including stable coins, debt tokens, share tokens, liquidity tokens, asset tokens, etc.} which can be invoked by regular users. The attacker manipulates the transfer amount under specific conditions. The mathematical models of victim contracts vary across different protocols.

Specifically, smart contracts with lending, leveraging, or yield-earning functions are more vulnerable to flash loan attacks. This susceptibility arises from the involvement of token transfers outside these contract types. Basic DeFi lego contracts, such as stable coins, Uniswap, and Curve.fi, are unlikely to be targeted in flash loan attacks due to their fundamental role as building blocks and infrastructure of nearly all DeFi protocols. These contracts have undergone exclusive audits and have been in use for several years.

Given a set of target smart contracts, {\gen} helps users of {\name} in selecting a set of action candidates likely to be involved in an attack vector.

\subsection{Selecting Action Candidates from Contract Application Binary Interfaces (ABIs)}
\label{subsec:abi}
The Application Binary Interface (ABI) serves as an interpreter enabling communication with the EVM bytecode. For all deployed and verified smart contracts on the blockchain, their ABIs are made public to facilitate user interactions by allowing them to call functions and engage with the contracts. An ABI typically comprises (public or external) function names, argument names/types, function state mutability, and return types.

During the process of selecting action candidates, certain functions can be safely ignored for the following reasons:
\begin{itemize}
    \item Functions with the view or pure mutability can be excluded as they do not modify the state of contracts.
    \item Functions that can only be invoked by privileged users, such as \emph{transferOwnership}, \emph{changeImplementation}, and \emph{changeAdmin}, can also be disregarded since they are unlikely to be accessed by regular users\footnote{Previous works~\cite{ghaleb2023achecker, liu2022finding} have extensively researched access control vulnerabilities, which are outside the scope of this work.}. This step can also be accomplished by conducting a lightweight static analysis on the source code of the contracts. 
    \item Token permission management functions/parameters, such as the \emph{approve} function or parameters like \textit{deadline} and \textit{amountOutMin}, are excluded. These functions/parameters solely control whether a transaction will be reverted or not and do not affect the functional behaviors of contracts. To simplify the search process of {\name}, these permissions are assumed to be granted maximally.
\end{itemize}

\subsection{Learning Special Parameters from Transaction History}
After selecting a set of action candidates from contracts' ABIs, some non-integer parameters(eg. bytes, string, address, array, enum) can still be unknown. {\gen} collects past transactions of the target contracts using TrueBlocks~\cite{trueblocks} and Blockchain Explorers(\cite{etherscan}, \cite{bscscan}, \cite{FTMScan}). Subsequently, {\gen} extracts function level trace data from these transactions using Phalcon\footnote{Phalcon is a transaction explorer that provides functional level trace data.}\cite{phalcon}, and utilizes this data to learn the special parameters from previous function calls made to the contract.

\subsection{Local Execution and Intra-dependency Analysis}
Following the selection of action candidates and determination of action parameters, each action candidate is executed in a forked blockchain (foundry~\cite{Foundry}) to verify its executability at a given block. An action candidate may not be executable due to various reasons: (1) the function is disabled by the owner or admin; (2) internal function calls to other contracts are disabled by the owners or admins of those contracts; (3) the function is not valid under current blockchain states. The inexecutability due to these reasons cannot be identified by static analysis of source code and can only be determined by executions. All such inexecutable functions are filtered out.

During this step, we focus solely on ensuring the executability of functions at the given block, without considering their profitability. {\gen} automatically collects storage read/write information during the execution of these functions and infers the Read-After-Write (RAW) dependencies\footnote{This RAW dependency information is also employed in {\name}'s initial data collection to expand the range of data points.} between different action candidates. An action A has a RAW dependency (or equivalently, is RAW dependent) on action B if the execution of action A reads the storage written by action B\footnote{It is important to note that this step excludes any tx.origin/msg.sender-related storage reads/writes, as such storage accesses do not alter the state of the protocol and are therefore unlikely to impact the functional behaviors of actions.}. From the RAW dependencies, it is possible to observe that certain functions behave independently, meaning they do not have any RAW dependencies on other functions, and other functions do not have any RAW dependencies on them either. Consequently, these independent functions can be safely ignored.

After analyzing ABIs, transaction history, and local executions, {\gen} generates a list of action candidates with only their integer arguments left undetermined. These action candidates are then input into {\name} for further synthesis.

 \subsection{\gen{} Algorithm}

 \setlength{\textfloatsep}{10pt}
 \begin{algorithm}[h!]
   \caption{\gen{} to identify candidate actions. 
   It takes a list of contract addresses $\contract$, a block height $\block$, and outputs a list of 
   candidate actions (functions), along with their RAW dependencies. }
   \label{alg:gen}
   \begin{algorithmic}[1]
   \Procedure{\gen{}}{$\contracts$, $\block$}
   \State \ $\functions  \leftarrow [\ ]$;
   \State \ $\traces  \leftarrow [\ ]$;
   \State \ \textbf{for each}\ $\A \in \contracts$; \label{algo3:line0}
   \State \ \ \ \ $\functions[\A] \leftarrow \textsc{AnalyzeABI}(\A)$
   \State \ \ \ \ $\tH \leftarrow  \textsc{CollectTransactions}(\A, \block) $; \label{algo3:line1}
   \State \ \ \ \ $\functions[\A].add( \textsc{Filter}(\tH) )$;  \label{algo3:line2}
   \State \ \ \ \ \textbf{for each}\ $a \in \functions[\A]$; 
   \State \ \ \ \ \ \ \ $\storageAccess[a] \leftarrow \textsc{Execute}(a, \block) $ \label{algo3:line2_2}
   \State \ $\dependentFunctions\leftarrow [\ ]$; \label{algo3:line3}
   \State \ \textbf{for each}\ $a \in \storageAccess.\keys$: \label{algo3:line4}
   \State \ \ \ \ \textbf{for each}\ $b \neq a \in \storageAccess.\keys$; 
   \State \ \ \ \ \ \ \ \ \textbf{if} \textsc{checkDependency}($\storageAccess[a]$, $\storageAccess[b]$)  \label{algo3:line7}
   \State \ \ \ \ \ \ \ \ \ \ \ $\dependentFunctions[a].add(b)$  \label{algo3:line8}
   \State \ \textbf{return} $\dependentFunctions$ \label{algo3:line9}
   \EndProcedure
   \end{algorithmic}
 \end{algorithm}

Algorithm \ref{alg:gen} describes the overall procedure \gen{} to identify candidate actions. 
\gen{} takes as inputs a list of contract addresses $\contracts$ that are used by a DeFi protocol and a block height $\block$. It then produces a list of functions $\dependentFunctions$ that constitute candidate actions. 

The procedure \gen{} first analyzes each contract's ABIs and identifies a list of non-privileged\footnote{Privileged functions refer to functions that can only be called by privileged users, e.g., owners of the contract, or within a certain range of blocks. Thus, these functions are not likely to be used by an attacker with an arbitrary address.} non-read-only functions with only integer arguments or no arguments. Then \gen{} inspects the blockchain state and collects the function level trace data for all historical transactions that occurred before the block height $\block$ for every contract listed in $\contracts$ using the sub-procedures \textsc{CollectTransactions} (line \ref{algo3:line1}). Using the trace data, we then leverage the sub-procedure \textsc{Filter} to identify non-integer arguments for other non-priviledged non-read-only functions of contract $\A$, and these functions, with only their integer parameters undetermined, are flagged as additional candidate actions (line \ref{algo3:line2}). We then execute these candidate actions locally to verify their executability via the sub-procedure \textsc{Execute}, while collecting their storage read/write information (line \ref{algo3:line2_2}). For each two functions $a$ and $b$, we check if $a$ is RAW dependent on $b$ via the subprocedure \textsc{checkDependency} (line \ref{algo3:line7}). If so, we add $b$ to the list of functions that $a$ depends on (line \ref{algo3:line8}). Finally, \gen{} returns the list of functions(keys of $\dependentFunctions$), along with their RAW dependencies (values of $\dependentFunctions$) (line \ref{algo3:line9}).

\subsection{Evaluation Results}
We conducted an evaluation of {\gen} and {\name} combined on $14$ historical flash loan attack benchmarks solved by {\name} alone. The detailed results are presented in Table 5 of main text. Notably, {\gen} and {\name} together successfully solved $11$ benchmarks. Surprisingly, for $6$ benchmarks, {\gen} discovered new attack vectors containing new actions that were not utilized by the original attackers. These new attack vectors exploit the same vulnerabilities and prove to be profitable as well, highlighting the effectiveness of {\gen} in identifying action candidates. 

However, there were 3 benchmarks where {\gen} failed to identify the correct set of action candidates. We discuss the reasons behind these failures below. For ElevenFi, there is a function \codeword{addLiquidity} which requires an argument \textit{amounts} of integer array of length 4. For AutoShark, a function \codeword{deposit} takes an address argument \textit{referrer}. For CheeseBank, a function \codeword{refresh} takes an argument \textit{symbols} which is a array of strings. After analyzing their past transactions, we find there are too many possible values for these non-integer arguments. Thus, we cannot learn the special parameters from transaction history. While experienced security analysts might manually identify these special parameters and include them in the action candidates, we view these instances as failures of {\gen}, as our goal is to automate the process as much as possible.

\section{Profit Calculation and Flash Loan Payment}
As discussed in Section 4 
, {\name} uses constant token
prices to calculate the profit of an attack vector. A positive profit
indicates flash loan vulnerabilities inside the contracts. We notice two
possible concerns regarding this assumption.

First, the price of a token might be influenced by an attack vector; 
thus {\name}'s profit calculation might be inaccurate. As a result, there
might be false positives in the results. In fact, hackers typically prefer to keep their final profit in stable coins or popular blockchain-native tokens such as Ethereum, Bitcoin, or BNB. This preference arises from the fact that other derivative tokens associated with victim protocols become worthless after the hack.

Second, if we require all initial
balances of an attacker is from flash loans,
a positive profit then does not necessarily 
mean the ability to pay back flash loan, since the final 
balances could not be strictly larger than the initial balances.
In historical flash loan attacks, the attackers usually choose to convert 
some of their profit to what's being flash loaned, for paying 
back the flash loan.

We designed an experiment to validate our results and justify 
our assumption. A summary of the experiment is shown in Table~\ref{tab:SupRepay}. For each benchmark, we select the attack vector with 
the best profit returned by {\name} under all different settings.
We record their final balances under "\textbf{Results of {\name}}" column. 
We observe for 8 benchmarks, their final balance can 
already cover the initial balance; thus the profit calculation is valid 
here, since it directly represents the increase of the attacker's balances. 

Apart from two security challenges(Puppet, PuppetV2), we tried to manually 
implement DEX swaps after the attack vectors, to make sure final 
balances are strictly larger than initial balances. Practically, 
we use 1inch~\cite{1inch} and Curve.Finance~\cite{curvefi, egorov2019stableswap} for
Ethereum benchmarks, and Pancake Swap~\cite{pancake} for Binance benchmarks. 
Results after swapping are listed under "\textbf{Results after Swapping}" column.
For $6$ benchmarks, we successfully implement DEX swaps so that their final 
balances are strictly larger than initial balances. 
We observe the profit calculation does not change a lot 
after swapping, which represents the attack vectors found by {\name} 
do yield positive profit and indicate flash loan vulnerabilities.

Overall, among $16$ historical flash loan attack benchamrks, 
{\name} synthesized attack vectors with positive profit for 
$14$ cases. And $13$ of them are valid even under the 
consideration of flash loan payment. 

\begin{table*}[t]
    \caption{Summary of Repaying Flash Loan(\textbf{Legends:} 
    ``\textbf{Initial Balance}'' represents initial token balances at the start of {\name}.  
    ``\textbf{Profit}'' represents adjusted calculated using constant token prices. 
    ``\textbf{Pay Back?}'' represents whether final token balances are strictly greater than initial balances.)}
    \label{tab:SupRepay}
    \centering
    
    \begin{tabular}{|l|l|l|l|l|l|l|l|}
    \hline
    \rowcolor[rgb]{ .851,  .851,  .851}
    benchmark    & Initial Balance                                                         & Results of {\name}                                                                                & Profit & Pay Back?   & \multicolumn{1}{l|}{Result after Swapping}                                            & \multicolumn{1}{l|}{Profit} & \multicolumn{1}{l|}{Pay Back?} \\ \hline
    bZx1         & \begin{tabular}[c]{@{}l@{}}112 WBTC \\ 4500 ETH\end{tabular}            & \begin{tabular}[c]{@{}l@{}}59 WBTC \\ 9009 ETH\end{tabular}                                              & 2438            & \xmark*  &  /                                                                                           &   /                                 &  /                          \\ \hline
    Harvest\_USDT & \begin{tabular}[c]{@{}l@{}}50000000 USDT \\ 20000000 USDC\end{tabular} & \begin{tabular}[c]{@{}l@{}}52635352 USDT  \\ 17502461 USDC\end{tabular}                                  & 137813          & \xmark   & \multicolumn{1}{l|}{\begin{tabular}[c]{@{}l@{}}50000000 USDT  \\ 20139106 USDC\end{tabular}} & \multicolumn{1}{l|}{139106}          & \multicolumn{1}{l|}{\cmark} \\ \hline
    Harvest\_USDC & \begin{tabular}[c]{@{}l@{}}18308555 USDT \\ 50000000 USDC\end{tabular} & \begin{tabular}[c]{@{}l@{}}15106415 USDT \\ 53312192 USDC\end{tabular}                                   & 110051          & \xmark   & \multicolumn{1}{l|}{\begin{tabular}[c]{@{}l@{}}18415434 USDT  \\ 50000000 USDC\end{tabular}} & \multicolumn{1}{l|}{106878}          & \multicolumn{1}{l|}{\cmark} \\ \hline
    Warp         & \begin{tabular}[c]{@{}l@{}}500000 WETH\\ 5000000 DAI\end{tabular}       & \begin{tabular}[c]{@{}l@{}}3917914 USDC\\ 493733 WETH\\ 8472025 DAI\end{tabular}                         & 3368718         & \xmark   & \multicolumn{1}{l|}{\begin{tabular}[c]{@{}l@{}}504549 WETH \\ 5000000 DAI\end{tabular}}      & \multicolumn{1}{l|}{2918865}         & \multicolumn{1}{l|}{\cmark} \\ \hline
    ValueDeFi    & \begin{tabular}[c]{@{}l@{}}116000000 DAI \\ 100000000 USDT\end{tabular} & \begin{tabular}[c]{@{}l@{}}102223937 DAI\\ 69829957 USDT  \\ 32023919 USDC  \\ 20300381 CRV\end{tabular} & 8378194         & \xmark   & \multicolumn{1}{l|}{\begin{tabular}[c]{@{}l@{}}116000000 DAI\\ 108225064 USDT\end{tabular}}  & \multicolumn{1}{l|}{8225064}         & \multicolumn{1}{l|}{\cmark} \\ \hline
    Yearn        & \begin{tabular}[c]{@{}l@{}}130000000 DAI \\ 134000000 USDC\end{tabular} &  /                                                                                                        & /                &  /        &  /                                                                                            &  /                                    & /                            \\ \hline
    Eminence     & 15000000 DAI                                                            & 1601468 DAI                                                                                              & 1601468         & \cmark   &  /                                                                                            &  /                                    & /                            \\ \hline
    CheeseBank   & 21000 ETH                                                               & \begin{tabular}[c]{@{}l@{}}20550 ETH  \\ 1435304 USDC  \\ 710427 USDT \\ 0 DAI\end{tabular}              & 1946291         & \xmark   & \multicolumn{1}{l|}{25493 ETH}                                                               & \multicolumn{1}{l|}{1991297}         & \multicolumn{1}{l|}{\cmark} \\ \hline
    InverseFi    & 27000 WBTC                                                              &  /                                                                                                        & /                &  /        &  /                                                                                            &  /                                    & /                            \\ \hline
    ElevenFi     & 130001 BUSD                                                             & 259742 BUSD                                                                                              & 129741          & \cmark   &  /                                                                                            &  /                                    & /                            \\ \hline
    bEarnFi      & 7804239 BUSD                                                            & 7818022 BUSD                                                                                             & 13783           & \cmark   &  /                                                                                            &  /                                    & /                            \\ \hline
    ApeRocket    & 1615000 CAKE                                                            & \begin{tabular}[c]{@{}l@{}}16849 WBNB\\ 1276757 CAKE\end{tabular}                                        & 1333            & \cmark   & \multicolumn{1}{l|}{1553 WBNB 1615000 CAKE}                                                  & \multicolumn{1}{l|}{1553}            & \multicolumn{1}{l|}{\cmark} \\ \hline
    AutoShark    & 100001 WBNB                                                             & 101373.63 WBNB                                                                                           & 1372            & \cmark   &  /                                                                                            &  /                                    & /                            \\ \hline
    Novo         & 2000 WBNB                                                               & 25084 WBNB                                                                                               & 23084           & \cmark   &  /                                                                                            &  /                                    & /                            \\ \hline
    Wdoge        & 3000 WBNB                                                               & 3078 WBNB                                                                                                & 78              & \cmark   &  /                                                                                            &  /                                    & /                            \\ \hline
    OneRing      & 150000000 USDC                                                          & 151893085 USDC                                                                                           & 1893085         & \cmark   &  /                                                                                            &  /                                    & /                            \\ \hline
    Puppet       & \begin{tabular}[c]{@{}l@{}}1000 DVT \\ 25 ETH\end{tabular}              & 100000 DVT  15 ETH                                                                                       & 89000           & \xmark** &  /                                                                                            &  /                                    & /                            \\ \hline
    PuppetV2     & \begin{tabular}[c]{@{}l@{}}10000 DVT \\ 20 ETH\end{tabular}             & \begin{tabular}[c]{@{}l@{}}886996 DVT\\ 1 ETH\\ 0 WETH\end{tabular}                                      & 857996          & \xmark** &  /                                                                                            &  /                                    & /                            \\ \hline
    \end{tabular}
    \flushleft
    *: bZx1 attack is the earliest flash loan attack (ETH block 9484688) among all benchmarks, 
    and we fail to find DEX with enough liquidity to perform the swap in the same transaction. 
    In history, after the exploit, bZx1 attacker purchased WBTC in small amounts over two days on DEX, 
    to repay the WBTC borrowed.   

    **: Unlike other benchmarks, Puppet and PuppetV2 are community security challenges instead of 
    flash loan attacks in history. We don't consider paying back flash loan in these two cases. 

\end{table*}

\section{{\name} Results under Different Settings}
We applied {\name} to $18$ benchmarks, we set a time limit of 1 hour for initial data
collection and another time limit of 2(resp., 3) hours for the rest of the synthesis procedure. 
In Section 7
, due to page limit, we are
only able to show a summary of our experiment results. 
Here we show all the results of our experiment 
under different settings and for all benchmarks.
The profit of best attack vectors found by {\name} under different settings is listed in 
Table~\ref{tab:SupProfit}. The synthesis time is listed in Table~\ref{tab:SupTime}.
The initial data collection time is listed in Table~\ref{tab:SupIC}. 

Note in Section 7
, we add synthesis time and initial data collection time
together to get the total time, while Table~\ref{tab:SupProfit} and Table~\ref{tab:SupTime}
list them separately.

\begin{table*}[thbp]
    \caption{Profit Summary under Different Settings
    (\textbf{N} and \textbf{N+X} represent running {\name} with $N$ initial data points for each action, 
    with CEGDC disabled and enabled respectively.  )}
    \addtolength{\tabcolsep}{-2.5pt}
    \centering

\begin{tabular}{|l|l|llllllll|} 
    \hline
    \rowcolor[rgb]{ .851,  .851,  .851}
    \multicolumn{2}{|l|}{}                     & \multicolumn{8}{c|}{FlashSyn-poly}                   \\ \hline
    \rowcolor[rgb]{ .851,  .851,  .851}
    benchmark             & historyProfit          & 200     & 200+X   & 500     & 500+X   & 1000    & 1000+X  & 2000    & 2000+X  \\ \hline
    bZx1                  & 1194                   & 2228    & 2443    & 2393    & 2336    & 2372    & 2446    & 2339    & 2392    \\
    Harvest\_USDT         & 338448                 & 117137  & 121132  & 136993  & 119232  & 137606  & 120713  & 137813  & 110139  \\
    Harvest\_USDC         & 307416                 & 53890   & 73021   & 88922   & 79355   & 90186   & 75377   & 59614   & 59614   \\
    Warp                  & 1693523                & 2778266 & 2969630 & 3368718 & 3368718 & 3290458 & 3290458 & 2773345 & 2773345 \\
    ValueDeFi             & 8618002                & 5512992 & 5512992 & 6919868 & 8088716 & 4831255 & 8307800 & 2562183 & 8378194 \\
    Yearn                 & 56924                  & /       & /       & /       & /       & /       & /       & /       & /       \\
    Eminence              & 1674278                & 1524534 & 1601468 & 1116830 & 1387119 & 1258248 & 1121077 & 585029  & 1507174 \\
    CheeseBank            & 3270347                & 1675958 & 1675958 & 1675958 & 1675958 & 1675958 & 1675958 & 1675958 & 1946291 \\
    InverseFi             & 2515606                & /       & /       & /       & /       & /       & /       & /       & /       \\
    ElevenFi              & 129741                 & 118025  & 129741  & 129741  & 129741  & 104279  & 105149  & 129741  & 129658  \\
    bEarnFi               & 18077                  & 13783   & 13730   & 13783   & 13756   & 13783   & 13715   & 13783   & 13770   \\
    ApeRocket             & 1345                   & 704     & 704     & /       & 1323    & 700     & 700     & 1333    & 1333    \\
    AutoShark             & 1381                   & /       & /       & 1346    & 1346    & /       & /       & /       & 1372    \\
    Novo                  & 24857                  & 14907   & 14907   & 17146   & 17146   & 14835   & 14835   & 14835   & 20210   \\
    Wdoge                 & 78                     & 75      & 75      & 75      & 75      & 75      & 75      & 75      & 75      \\
    OneRing               & 1534752                & 1673425 & 1673425 & 1814877 & 1814877 & 628440  & 628440  & 590506  & 1814882 \\
    Puppet                & 89000                  & 85093   & 85092   & 84095   & 84096   & 89000   & 78200   & 89000   & 89000   \\
    PuppetV2              & 953100                 & 760543  & 835148  & 326400  & 400830  & 647894  & 857996  & 400965  & 747799  \\ \hline
    \multicolumn{2}{|r|}{Solved:}                  & 15      & 15      & 15      & 16      & 15      & 15      & 15      & 16      \\ 
    \multicolumn{2}{|r|}{Avg. Normalized. Profit:} & 0.793   & 0.829   & 0.846   & 0.922   & 0.762   & 0.786   & 0.717   & 0.945   \\ \hline 
    \end{tabular}

\vspace{5mm}

\begin{tabular}{|l|l|llllllll|} 
    \hline
    \rowcolor[rgb]{ .851,  .851,  .851}
    \multicolumn{2}{|l|}{}                     & \multicolumn{8}{c|}{FlashSyn-inter}                                           \\ \hline
    \rowcolor[rgb]{ .851,  .851,  .851}
    benchmark             & historyProfit          & 200     & 200+X   & 500     & 500+X   & 1000    & 1000+X  & 2000    & 2000+X  \\ \hline
    bZx1                  & 1194                   & 2018    & 2383    & 2364    & 2364    & 2231    & 2294    & 2301    & 2302    \\
    Harvest\_USDT         & 338448                 & 739     & 10759   & 6981    & 13944   & 12043   & 72597   & 1544    & 86798   \\
    Harvest\_USDC         & 307416                 & 506     & 14338   & 65670   & 506     & 505     & 2125    & 504     & 110051  \\
    Warp                  & 1693523                & /       & /       & 596570  & 596570  & /       & /       & /       & /       \\
    ValueDeFi             & 8618002                & 5415559 & 5415559 & 8348059 & 8348059 & 287732  & 3468019 & 6428341 & 6428341 \\
    Yearn                 & 56924                  & /       & /       & /       & /       & /       & /       & /       & /       \\
    Eminence              & 1674278                & /       & /       & /       & /       & /       & /       & /       & /       \\
    CheeseBank            & 3270347                & 659895  & 659895  & 952455  & 1068853 & 1045789 & 1339468 & 1101547 & 1101547 \\
    InverseFi             & 2515606                & /       & /       & /       & /       & /       & /       & /       & /       \\
    ElevenFi              & 129741                 & 57510   & 57510   & 65090   & 86687   & 94456   & 94456   & 85811   & 85811   \\
    bEarnFi               & 18077                  & 11448   & 11448   & 12695   & 11787   & 11576   & 12059   & 11335   & 12329   \\
    ApeRocket             & 1345                   & 1323    & 1323    & 1333    & 1333    & 1135    & 1135    & 1037    & 1037    \\
    AutoShark             & 1381                   & /       & /       & /       & /       & /       & /       & /       & /       \\
    Novo                  & 24857                  & 14317   & 14317   & 12196   & 13891   & 14065   & 14065   & 23084   & 23084   \\
    Wdoge                 & 78                     & 74      & 74      & 78      & 78      & 75      & 75      & 75      & 75      \\
    OneRing               & 1534752                & 1893085 & 1893085 & 1830059 & 1891589 & 1888510 & 1888510 & 1814877 & 1942188 \\
    Puppet                & 89000                  & 81576   & 82560   & 78818   & 78821   & 81172   & 81172   & 87266   & 87266   \\
    PuppetV2              & 953100                 & 344943  & 224177  & 459980  & 459980  & 397256  & 397256  & 362540  & 362541  \\ \hline
    \multicolumn{2}{|r|}{Solved:}                  & 13      & 13      & 14      & 14      & 13      & 13      & 13      & 13      \\ 
    \multicolumn{2}{|r|}{Avg. Normalized. Profit:} & 0.539   & 0.555   & 0.630   & 0.634   & 0.535   & 0.580   & 0.594   & 0.641   \\ \hline 
    \end{tabular}

\label{tab:SupProfit}
\end{table*}

\begin{table*}[thbp]
    \caption{{\name}'s Synthesis Time Summary under Different Settings. }
    \addtolength{\tabcolsep}{-2.5pt}
    \centering
    \begin{tabular}{|l|llllllll|llllllll|}
        \hline
        \rowcolor[rgb]{ .851,  .851,  .851}
        Time(s)       & \multicolumn{8}{c|}{FlashSyn-poly}                              & \multicolumn{8}{c|}{FlashSyn-inter}                               \\ \hline
        \rowcolor[rgb]{ .851,  .851,  .851}
        benchmark     & 200  & 200+X & 500  & 500+X & 1000 & 1000+X & 2000 & 2000+X & 200   & 200+X & 500   & 500+X & 1000  & 1000+X & 2000  & 2000+X \\ \hline
        bZx1          & 111  & 379   & 115  & 271   & 131  & 454    & 119  & 321    & 159   & 357   & 146   & 327   & 160   & 1634   & 129   & 340    \\
        Harvest\_USDT & 144  & 636   & 295  & 1176  & 376  & 621    & 320  & 548    & 3981  & 7940  & 5985  & 8380  & 5363  & 7867   & 5616  & 7457   \\
        Harvest\_USDC & 188  & 554   & 365  & 991   & 350  & 451    & 420  & 561    & 3662  & 7611  & 5845  & 8400  & 5526  & 7845   & 5476  & 8233   \\
        Warp          & 593  & 1484  & 1319 & 1542  & 1175 & 1428   & 965  & 1069   & /     & /     & 3320  & 5446  & /     & /      & /     & /      \\
        ValueDeFi     & 1098 & 1079  & 826  & 5339  & 927  & 3669   & 1145 & 4402   & 10800 & 10800 & 10800 & 10800 & 10800 & 10800  & 10800 & 10800  \\
        Yearn         & /    & /     & /    & /     & /    & /      & /    & /      & /     & /     & /     & /     & /     & /      & /     & /      \\
        Eminence      & 468  & 1525  & 461  & 1251  & 1025 & 1209   & 460  & 1094   & /     & /     & /     & /     & /     & /      & /     & /      \\
        CheeseBank    & 4311 & 4080  & 4811 & 5586  & 4648 & 6379   & 5022 & 4249   & 9840  & 10800 & 10800 & 10800 & 10800 & 10800  & 10800 & 10800  \\
        InverseFi     & /    & /     & /    & /     & /    & /      & /    & /      & /     & /     & /     & /     & /     & /      & /     & /      \\
        ElevenFi      & 131  & 298   & 156  & 300   & 139  & 1352   & 132  & 288    & 457   & 875   & 463   & 834   & 374   & 2443   & 381   & 777    \\
        bEarnFi       & 123  & 359   & 118  & 601   & 155  & 731    & 162  & 439    & 352   & 673   & 569   & 950   & 664   & 666    & 517   & 657    \\
        ApeRocket     & 230  & 409   & /    & 610   & 224  & 529    & 416  & 652    & 2057  & 3132  & 2342  & 3724  & 1921  & 3047   & 1920  & 3157   \\
        AutoShark     & / & /  & 4485 & 5762  & / & /   & /  & 5348   & /   & /   & /   & /   & /   & /    & /   & /    \\
        Novo          & 239  & 420   & 524  & 721   & 330  & 512    & 383  & 630    & 468   & 740   & 796   & 879   & 482   & 848    & 657   & 789    \\
        Wdoge         & 89   & 203   & 170  & 299   & 145  & 377    & 136  & 237    & 117   & 263   & 226   & 427   & 106   & 385    & 217   & 254    \\
        OneRing       & 92   & 192   & 183  & 378   & 120  & 309    & 159  & 526    & 110   & 259   & 194   & 485   & 97    & 383    & 169   & 308    \\
        Puppet        & 36   & 68    & 37   & 65    & 21   & 41     & 32   & 83     & 86    & 197   & 72    & 158   & 89    & 146    & 65    & 118    \\
        PuppetV2      & 45   & 133   & 48   & 132   & 35   & 75     & 37   & 103    & 233   & 484   & 196   & 456   & 218   & 437    & 220   & 497    \\ \hline
        Avg.Time(s)   & 527  & 788   & 928  & 1564  & 653  & 1209   & 661  & 1284   & 2486  & 3395  & 2982  & 3719  & 2815  & 3639   & 2844  & 3399   \\ \hline
        \end{tabular}
\label{tab:SupTime}
\end{table*}

\begin{table*}[thbp]
    \caption{{\name}'s Initial Data Collection Time Summary under Different Settings. }
    \addtolength{\tabcolsep}{-2.5pt}
    \centering
    
        \begin{tabular}{|l|llll|}
            \hline
            \rowcolor[rgb]{ .851,  .851,  .851}
        Initial Data Collection(s) &         &         &          &          \\ \hline
        \rowcolor[rgb]{ .851,  .851,  .851}
        benchmark                  & 200     & 500     & 1000     & 2000     \\ \hline
        bZx1                       & 49      & 73      & 91       & 101      \\
        Harvest\_USDT              & 37      & 49      & 74       & 122      \\
        Harvest\_USDC              & 62      & 76      & 68       & 116      \\
        Warp                       & 33      & 45      & 60       & 95       \\
        ValueDeFi                  & 80      & 113     & 172      & 289      \\
        Yearn                      & 168     & 189     & 217      & 299      \\
        Eminence                   & 56      & 63      & 73       & 97       \\
        CheeseBank                 & 91      & 94      & 107      & 142      \\
        InverseFi                  & 52      & 62      & 91       & 152      \\
        ElevenFi                   & 69      & 72      & 95       & 121      \\
        bEarnFi                    & 15      & 15      & 23       & 31       \\
        ApeRocket                  & 29      & 36      & 48       & 81       \\
        AutoShark                  & 99      & 97      & 123      & 136      \\
        Novo                       & 36      & 42      & 51       & 72       \\
        Wdoge                      & 10      & 14      & 21       & 35       \\
        OneRing                    & 17      & 22      & 36       & 59       \\
        Puppet                     & 220     & 633     & 686      & 1120     \\
        PuppetV2                   & 773     & 1482    & 1218     & 2338     \\ \hline
        Avg. Time:                 & 105     & 176     & 181      & 300     \\ \hline
            
    \end{tabular}

\label{tab:SupIC}
\end{table*}

\begin{table*}[thbp]
    \caption{Number of Total Data Points under Different Settings(
        \textbf{N(M)} represents the number of data points is \textbf{N} after \textbf{M} rounds of CEGDC. 
    )}
    \centering

    \begin{tabular}{|l|llllllll|}
        \hline
        \rowcolor[rgb]{ .851,  .851,  .851}
        \multicolumn{1}{|l|}{}                      & \multicolumn{8}{c|}{FlashSyn-poly} \\ \hline
        \rowcolor[rgb]{ .851,  .851,  .851}
        benchmark         & 200  & 200+X & 500  & 500+X & 1000 & 1000+X & 2000  & 2000+X \\ \hline
        bZx1              & 600  & 1788  & 1500 & 2273  & 3000 & 3892   & 5192  & 5849   \\
        Harvest\_USDT     & 800  & 2002  & 2000 & 3704  & 4000 & 5330   & 8000  & 9325   \\
        Harvest\_USDC     & 800  & 3004  & 2000 & 3561  & 4000 & 4695   & 8000  & 8912   \\
        Warp              & 600  & 604   & 1500 & 1500  & 3000 & 3001   & 6000  & 6000   \\
        ValueDeFi         & 1200 & 1200  & 3000 & 11138 & 6000 & 13060  & 12000 & 19975  \\
        Yearn             & /    & /     & /    & /     & /    & /      & /     & /      \\
        Eminence          & 800  & 1525  & 2000 & 2663  & 4000 & 4659   & 8000  & 8780   \\
        CheeseBank        & 464  & 464   & 942  & 715   & 1534 & 1770   & 2679  & 2937   \\
        InverseFi         & /    & /     & /    & /     & /    & /      & /     & /      \\
        ElevenFi          & 400  & 460   & 1000 & 1077  & 2000 & 2078   & 4000  & 4070   \\
        bEarnFi           & 400  & 1444  & 1000 & 2066  & 2000 & 2744   & 4000  & 4854   \\
        ApeRocket         & 600  & 630   & 1500 & 1739  & 3000 & 3223   & 6000  & 6402   \\
        AutoShark         & 478  & 478   & 1068 & 1068  & 1645 & 1645   & 2753  & 2753   \\
        Novo              & 400  & 578   & 1000 & 1110  & 2000 & 2072   & 4000  & 4164   \\
        Wdoge             & 200  & 201   & 500  & 501   & 1000 & 1001   & 2000  & 2001   \\
        OneRing           & 400  & 662   & 1000 & 1386  & 2000 & 2188   & 4000  & 4710   \\
        Puppet            & 600  & 712   & 1500 & 1585  & 3000 & 3111   & 6000  & 6301   \\
        PuppetV2          & 600  & 914   & 1396 & 1936  & 2534 & 2664   & 4491  & 4836   \\ \hline
        Avg. data points: & 584  & 1042  & 1432 & 2376  & 2795 & 3571   & 5445  & 6367   \\ \hline
    \end{tabular}

    \vspace{5mm}

    \begin{tabular}{|l|llllllll|}
        \hline
        \rowcolor[rgb]{ .851,  .851,  .851}
        \multicolumn{1}{|l|}{}            & \multicolumn{8}{c|}{FlashSyn-inter}   \\ \hline
        \rowcolor[rgb]{ .851,  .851,  .851}
        benchmark         & 200  & 200+X & 500  & 500+X & 1000 & 1000+X & 2000  & 2000+X \\ \hline
        bZx1              & 600  & 1965  & 1500 & 3028  & 3000 & 4413   & 5192  & 6373   \\
        Harvest\_USDT     & 800  & 3367  & 2000 & 4811  & 4000 & 6567   & 8000  & 10289  \\
        Harvest\_USDC     & 800  & 3687  & 2000 & 5240  & 4000 & 7231   & 8000  & 10914  \\
        Warp              & 600  & 600   & 1500 & 1500  & 3000 & 3000   & 6000  & 6000   \\
        ValueDeFi         & 1200 & 1200  & 3000 & 5974  & 6000 & 9281   & 12000 & 15758  \\
        Yearn             & /    & /     & /    & /     & /    & /      & /     & /      \\
        Eminence          & 800  & 923   & 2000 & 2080  & 4000 & 4109   & 8000  & 8104   \\
        CheeseBank        & 464  & 464   & 942  & 1099  & 1534 & 1635   & 2679  & 2715   \\
        InverseFi         & /    & /     & /    & /     & /    & /      & /     & /      \\
        ElevenFi          & 400  & 759   & 1000 & 1341  & 2000 & 2363   & 4000  & 4326   \\
        bEarnFi           & 400  & 1762  & 1000 & 2848  & 2000 & 2696   & 4000  & 4652   \\
        ApeRocket         & 600  & 1659  & 1500 & 2426  & 3000 & 3309   & 6000  & 6235   \\
        AutoShark         & 478  & 478   & 1068 & 1500  & 1645 & 1645   & 2753  & 2753   \\
        Novo              & 400  & 412   & 1000 & 1035  & 2000 & 2028   & 4000  & 4031   \\
        Wdoge             & 200  & 431   & 500  & 654   & 1000 & 1137   & 2000  & 2080   \\
        OneRing           & 400  & 656   & 1000 & 1284  & 2000 & 2218   & 4000  & 4218   \\
        Puppet            & 600  & 1244  & 1500 & 2012  & 3000 & 3509   & 6000  & 6452   \\
        PuppetV2          & 600  & 1806  & 1396 & 2369  & 2534 & 3348   & 4491  & 5061   \\ \hline
        Avg. data points: & 584  & 1338  & 1432 & 2450  & 2795 & 3656   & 5445  & 6248   \\ \hline
    \end{tabular}

\label{tab:SupDataPoint}
\end{table*}

\end{document}